\documentclass[useAMS,usenatbib,usenatbib,usegraphicx]{mn2e}

\usepackage{natbib}
\usepackage{latexsym}

\bibpunct{(}{)}{;}{a}{}{,}

\begin{document} 

\title{Observable Signatures of the low-z Circum-Galactic and Inter-Galactic Medium : UV Line Emission in Simulations}

\author[S. Frank et al.]{S. Frank$^{1}$\thanks{E-mail: sfrank@oamp.fr}, Y. Rasera$^{2}$, D. Vibert$^{1}$,
 \newauthor
  B. Milliard$^{1}$, A. Popping$^{1,4}$, J. Blaizot$^{3}$, S. Courty$^{3}$, J.-M. Deharveng$^{1}$,
 \newauthor
 C. P\'eroux$^{1}$, R. Teyssier$^{2}$, C. D. Martin $^{5}$ \\
$^{1}$Laboratoire d'Astrophysique de Marseille, OAMP, Universit\'e Aix-Marseille \&{} CNRS\\
38 rue Fr\'ed\'eric Joliot Curie, 13388 Marseille cedex 13, France\\
$^{2}$ Observatoire de Paris, Universit\'e Paris Diderot; 5 Place Jules Janssen,\\ 92190 Meudon, France), CNRS\\
$^{3}$ Universit\'e de Lyon, Lyon, F-69003, France ; Universit\'e Lyon 1, Observatoire de Lyon, 9 avenue Charles Andr\'e, Saint-Genis Laval, F-69230, France ; CNRS, UMR 5574, Centre de Recherche Astrophysique de Lyon ; Ecole Normale Sup\'erieure de Lyon, Lyon, F-69007, France\\
$^{4}$ International Centre for Radio Astronomy Research, The University of Western Australia, M468,\\35 Stirling Highway, Crawley, WA 6009, Australia\\
$^{5}$ California Institute of Technology, Cahill Center for Astrophysics, 1216 E. California Blvd.,\\Pasadena, CA 91125 USA\\
}

\date{Accepted : 2011 Novemeber 08. Received 2011 November 08; in original form 2011 September 14}

\pagerange{\pageref{firstpage}--\pageref{lastpage}} 

\maketitle

\label{firstpage}

\begin{abstract}
We present for the first time predictions for UV line emission of intergalactic and circumgalactic gas from Adaptive Mesh Resolution (AMR) Large Scale Structure (LSS) simulations at redshifts 0.3$<$z$<$1.2, with a specific emphasis on its observability with current and near-future UV instrumentation. In the three UV transitions of interest (Ly$\alpha$, OVI and CIV) there is a clear bimodality in the type of emitting objects : the overwhelming majority of the flux stems from discrete, compact sources, while a much larger fraction of the volume is filled by more tenuous gas. We characterise both object types with regard to their number densities, physical sizes and shapes, brightnesses and luminosities, velocity structures, masses, temperatures, ionisation states, and metal content. Degrading our AMR grids to characteristic resolutions offered by  available (such as FIREBall) or foreseeable instrumentation, allows us to assess which inferences can be drawn from currently possible observations, and to set foundations to prepare observing strategies for future missions. In general, the faint emission of the IGM and filamentary structure remains beyond the capabilities of instruments with only short duration exposure potential (i.e. stratospheric balloons), even for the most optimistic assumption for Ly$\alpha$, while the yet fainter metal line transitions (OVI and CIV) for these structures will actually remain challenging for long duration exposures (i.e. space-based telescopes), mostly due to their low metallicities pushing them more than three orders of magnitudes in brightness below the Ly$\alpha${} radiation. For the bright, circum-galactic medium (CGM) the situation is much more promising, and it is foreseeable that in the near future we will not only just dectect such sources, but the combination of all three lines in addition to velocity information will yield valuable insight into the physical processes at hand, illuminating (and discriminating between) important mechanisms during the formation of galaxies and their backreaction onto the IGM from whence they formed.          
\end{abstract}


\section{Introduction}\label{introduction}

Both inventories of the cosmic baryons budget \citep[]{fukugita1998, fukugita2004}, and cosmological simulations \citep[]{cen1999, dave1999, dave2001, cen2006} which predict the development of a warm-hot phase of the intergalactic medium (IGM) at low redshift 
have motivated an intensive search for baryons in a physical state 
not well explored yet (e.g. \citet{bregman2007, lehner2007, danforth2008, tripp2008, thom2008, danforth2010} and references therein). \\

Beyond the baryon accounting issue itself, this approach is providing new insights into important aspects of cosmic evolution, specifically about the low fraction of baryons (about 6 \%) that have made their way into stellar populations of galaxies, the  interdependence between the IGM and galaxy evolution, and  the crucial role of baryons circulating in and out galaxies.

The exploration of matter in the IGM at densities below the mean density is naturally  performed through quasar (or other bright background sources) absorption-line studies, and has a long history of spectacular successes. At low redshift, in the domain of tracers of the Warm-Hot intergalactic medium (WHIM), this is now culminating in observations with the Cosmic Origins Spectrograph (COS) on board of HST (e.g. \citet[]{narayanan2011, savage2011}.\\
   
Absorption line studies are, however, traditionally limited by a number of factors, such as the need for bright background sources, the difficulty of exploring perpendicular to a line of sight, and the difficulty of finding background sources for investigating a specific object. In addition any improvement in sensitivity requires an increase of collecting area. Emission line observations, as discussed by \citet[]{hogan1987}{} and \citet[]{gould1996}, have therefore been considered as an alternative approach. They are not affected by the issue of background sources and, depending on the observed solid angle, are less demanding than absorption techniques in terms of collecting area. In this vein, e.g. \citet[]{rauch2008} have performed a very long exposure with VLT in order to detect Lyman $\alpha$ emission from matter at the transition from ionized gas to neutral, self-shielded gas. \\
    
Exploiting a variety of simulations, predictions have been made for the possible signatures of structure formation and 'feedback processes' in terms of emission lines over a range of redshifts. Focusing on hydrogen as the naturally most abundant species, \citet[]{furlanetto2004, furlanetto2005, kollmeier2010} and references therein, predict Ly$\alpha${} emission from IGM gas at $0 \leq z \leq 5$, and discuss various mechanisms for its origin (reradiation of absorbed ionising photons from the metagalactic UV background; radiative cooling of shocks during the assembly of gravitationally bound objects; interaction of strong winds with the surrounding IGM). Furthermore, these authors also stress the difficulty to discriminate between these, in addition to the complication that galaxies themselves emit in Ly$\alpha$. \citet[]{bertone2010b} and \citet{bertone2010c} trace a large suite of metals in different ionisation stages through their SPH simulation in combination with state of the art gas cooling prescriptions to track their emission both in the soft-X rays as well as in the (restframe) UV over a similar redshift range as above(0$<$z$<$5). Of particular interest are their notions that the transitions in CIII, CIV, SiIV and OVI are bright enough ($>$1000 photons/(s cm$^2$ sr) to allow for a detection with current technology ('in the near future') in gas of moderate overdensities ($\delta \geq 100$) that has been enriched to metallicities Z$>0.1 Z_{solar}$. They caution, however, that hence these emission lines are biased tracers of the baryons, but still provide good tools to detect gas cooling onto or flowing out of galaxies.\\

In this paper, we follow a similar line of investigation for Ly$\alpha$, CIV and OVI emission over the redshift interval 0.3$<$z$<$1.2, but go one step further in the detailed effects of observability with current and/or near-future instrumentation in the UV. We clearly highlight the limitations posed by the detectors' spatial resolution, and stress the importance of taking the velocity structure of the emitting gas into account. We run source finding software in simulated data-cubes on grids resembling a typical instrumentational set-up, and are for the first time able to characterise potentially detectable structures with respect to their brightness, luminosities, masses, surface and volume densities, sizes and morphologies. There is a clear delineation (in Ly$\alpha$) between the very diffuse and extended filamentary structures tracing the true IGM, and 'haloes' surrounding the  bright and massive nodes of the Cosmic Web, which are more compact ($<$ 200 kpc/h in diameter) and much brighter than the filamentary sheets.\\
 
The paper is organised as follows. We briefly discuss the simulations in Section 2, highlighting the methods to calculate the gas emissivities in \ref{sec:gas_emission}, and add a description how to arrive at an observable data cube in \ref{sec:observed_cubes}. In section \ref{sec:general_features} we take a first look at these cubes in all three emission lines we are primarily interested in (Ly$\alpha$, CIV, and OVI), and discuss general features of the emission in terms of brightness distribution functions and spatial extent, while sections \ref{sec:extended_sources} and \ref{sec:compact_sources} provide a closer look at the extended, filamentary structures and the compact, bright sources, respectively. Before concluding in section \ref{sec:conclusions}, we discuss the implications of our findings in terms of observing strategies and optimal technical instrument specifications in section \ref{sec:observing_strategies}.\\
Throughout the paper we use a $\Lambda$CDM cosmology with the results of WMAP5 (see next section for detailed parameters).


\section{Simulations}
\subsection{Cosmological simulations}

The high-resolution cosmological simulation investigated in this article is part of the BINGO! simulation\footnote{BINGO! is a 4-year project of the ''Agence Nationale de la Recherche"(ANR), started in 2009. The consortium of three partner institutes (Marseille, Lyon, and Paris) specifically investigates both theoretically and observationally the physical processes of structure and galaxy formation with an emphasis on their interactions with the IGM and CGM.} suite. It was performed  at the CINES computational center with the RAMSES Adaptive Mesh Refinement (AMR) code (\cite{teyssier2002}). Initial conditions were set using the MPGRAFIC code (\cite{prunet2008}), a parallel implementation of GRAFIC (\cite{bertschinger2001}), and made use of the HORIZON project white noise\footnote{www.projet-horizon.fr}. They assume the following cosmological parameters, consistent with the results of WMAP5: $\Omega _{m}(0) = 0.26, \Omega _{\Lambda}(0) = 0.74, h = 0.719, \Omega _{b} = 0.044, \sigma _{8} = 0.796, n _{s} = 0.96$.

The comoving box length is 100~Mpc/h , the number of dark matter particles is $512^3$, the number of cells at the coarse level is $512^3$ as well, and the number of refinement levels is 7. As a consequence, the dark matter particle mass is $m_p=4.42\times 10^8$~M$_\odot/h$, and the spatial resolution at the highest refinement level is roughly constant and equal to $1.53$ physical kpc/$h$.

RAMSES computes the evolution of dark matter particles, gas cells, and star particles within the AMR hierarchy of grids. Particles are evolved using a PM method with a multigrid Poisson Solver. Euler equations for the gas are solved with a second-order unsplit Godunov method. We adopt the UV background from Haardt and Madau (1996) with a zero percent escape fraction from galaxies below the Lyman limit. For instance, the flux at z=0.75 and at 912 \AA{} is 11.75 photons cm$^{-2}$s$^{-1}$sr$^{-1}$ A$^{-1}$. The net cooling rate in the simulation is dependent on the density and metallicity of the gas, using a parametrization based on cooling curves calculated with Cloudy for a gas in ionization equilibrium with the adopted UV background.

It is worth noting that most of the work about UV emission lines of the WHIM/CGM was done with SPH (\cite{furlanetto2004,bertone2010b}) or fixed grid codes (\cite{cen2006}), whereas the modelisation of mildly-overdense shock-heated regions might be subject to caution with these Lagrangian or low-resolution codes.  While Smith et al. (2011) also use an AMR code (ENZO), they are focusing mostly on characterising the WHIM and OVI absorption rather than emission from CGM and IGM gas as in our case. 

Star formation and feedback follow the methodology described in Dubois and Teyssier(2008, 2010). Star formation occurs in cells denser than a density threshold of $n_{\rm H}  = 0.1$ at/cm$^3$, at a rate proportional to the gas density with an efficiency of 2\% per free-fall time. The effect of type II SN explosions is taken into account by releasing within a radius $r_{sn}=15$~kpc from each stellar particle, 10Myr after its formation, (i) 10\% of the mass formed, (ii) 10$^51$erg per SN event, half in the form of kinetic energy and half in the form of thermal energy, and (iii) metals, assuming a global yield of 0.1. Metals are then advected as a passive scalar enriching the ISM, but also the CGM and WHIM of interest here in this article. Finally, a polytropic equation of state of index $\gamma=5/3$ is used to provide pressure support to the dense gas ($n_{\rm H} > 0.1$ at/cm$^3$) and avoid artificial fragmentation.

\subsection{Gas emission mechanisms accounted for by the simulation (photoionized and collisionally excited gas)} \label{sec:gas_emission}

The estimate of gas line emissivities follows naturally from the simulations through the use of grids indexed by hydrogen density and temperature, assuming the same uniform ionizing background as adopted for the simulations and hence being self-consistent regarding this detail. We have used the c8.00 version of the spectral synthesis code Cloudy\footnote{http://www.nublado.org}, last described by Ferland et al. (1998), to compute the line emissivities of a plasma submitted to an incident continuum. The line emissivities of interest in this paper are Lya at 1216 \AA, OVI at 1032/1038 \AA and CIV at 1548/1551 \AA. Grids have been generated at the redshifts z=0.35, 0.76 and 1, and for each redshift, the incident continuum used in Cloudy is the same ultra-violet background used in the simulation at the same redshift. Practically, these specific redshift values have been adopted in order to place the redshifted lines in the unique  narrow spectral UV window at 2000 A accessible from stratospheric balloon-borne experiments such as FIREBall; in the following we will explore the redshift influence since our goal is to  understand gas emission over the redshift range 0.3 - 1 . \\

Grid points are computed for temperatures between $3 \times 10^3$ K and $3 \times 10^8$ K every $\Delta log(T/K)=0.01$ and for densities between $n_H=1.778\times10^{-8}$ and $n_H=1$ cm$^{-3}$ every $\Delta log(n_H/cm^{-3})=0.5$.  Two sets of grids are calculated, one assuming the plasma to be in photoionization equilibrium with the UV background and another one assuming collisional ionization equilibrium. The grids are computed at solar metallicity and the emissivities of each gas cells is rescaled according to their gas metallicity. We use the ``no induced processes'' of Cloudy (for an explanation for this choice see e.g. \citet{furlanetto2004}{} and the following treatment of the photon pumping). \\

\begin{figure*}
\includegraphics[width=168mm]{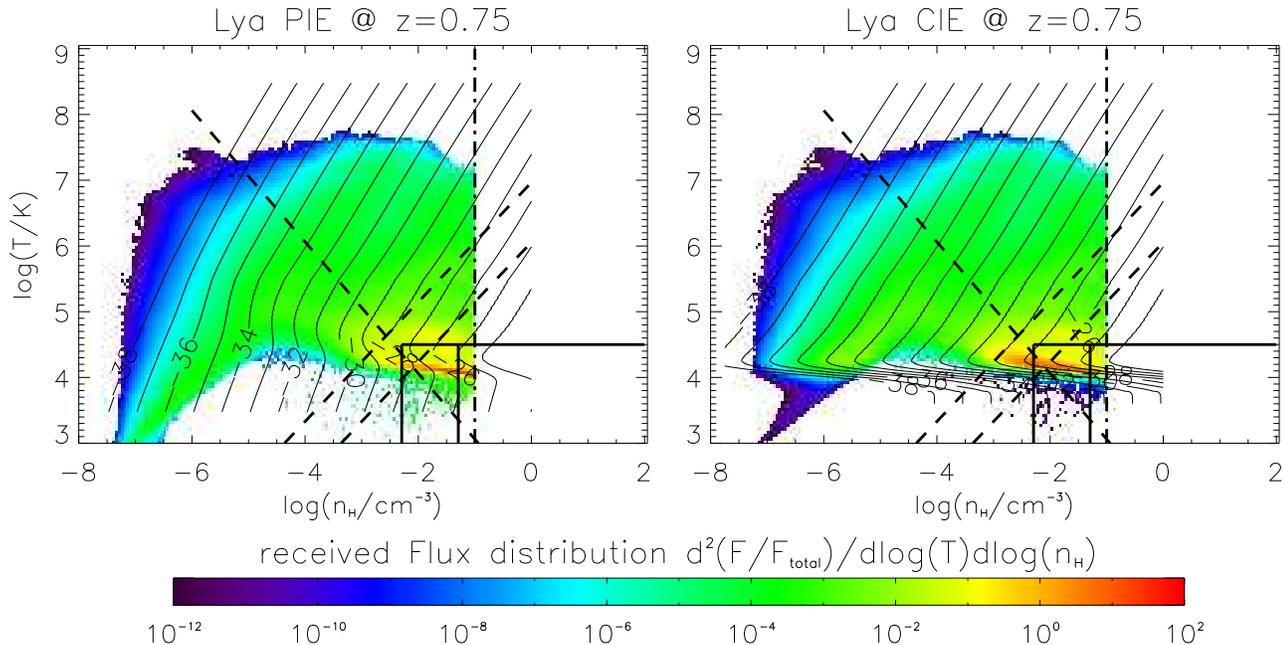}\caption{The relative flux distribution for the simulated 100 Mpc/h box in the log T - log n$_{H}${} bins (coloured phase-space diagram), and the Ly $\alpha${} volume emissivity (black solid lines, in units of erg/(s$\times cm^{3})$) as function of hydrogen density and gas temperature using only photo-ionisation equilibrium (PIE, left panel) or  collisional ionisation equilibrium (CIE,right panel) assumptions at a redshift of z=0.75. In addition, the location of the cuts used to delineate various scenarios for the self-shielding treatment is indicated by the dashed lines plus the two rectangular boxes on the lower right (for details, see text). It is clear that for Ly$\alpha${} emission the exact location of those cuts plays a vital role in determining the total emission, as gas with high density and T$\sim 10^{4}$ K dominates the emission. Note that in this and the following plots, the choices of redshifts stem from our restriction to a constant wavelength coverage over a well-defined UV regime of a putative instrument, in order to be realistically illustrative.}\label{lya_pie_emissivity}
\end{figure*}

\begin{figure*}
\includegraphics[width=168mm]{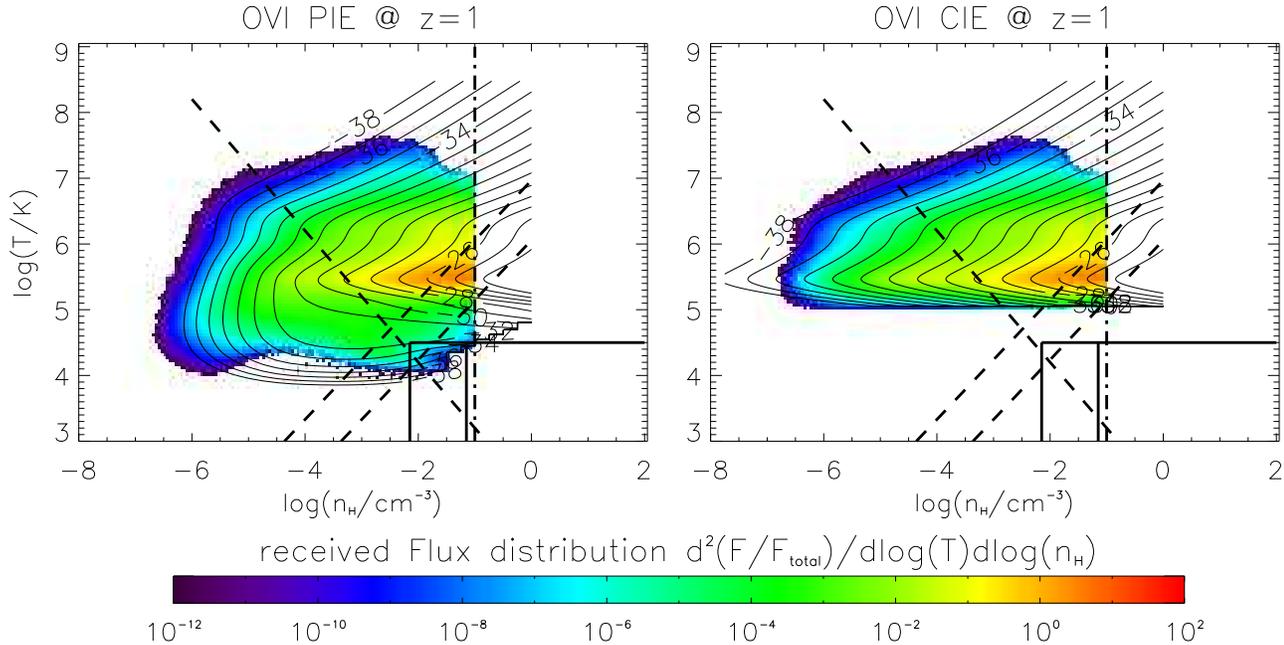}\caption{Same structure as in Figure \ref{lya_pie_emissivity}, now showing OVI at a redshift of z=1.0. Note how here (and in the following plot for CIV) the area of main emissivity shifts to much higher temperatures (above 10$^{5}${} K), and thus the exact location of the self-shielding cut does not play an important role anymore for determining the total flux.}
\label{ovi_pie_emissivity}
\end{figure*}

\begin{figure*}
\includegraphics[width=168mm]{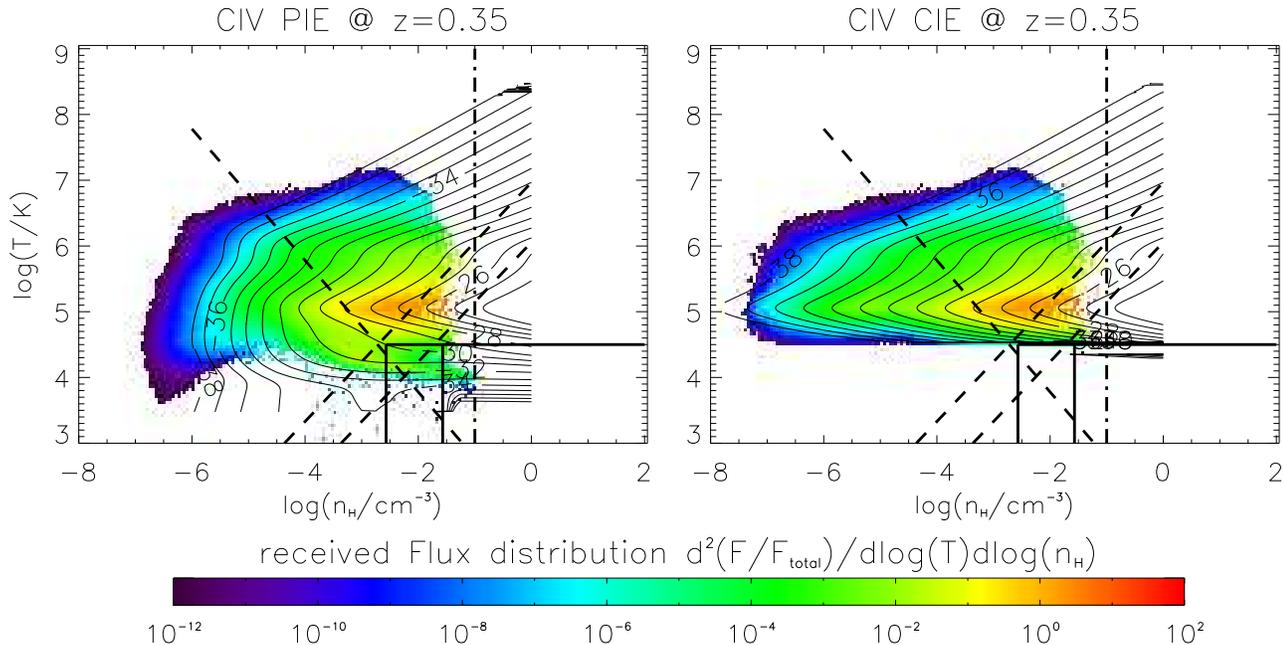}\caption{Same structure as in Figure \ref{lya_pie_emissivity}, now showing CIV at a redshift of z=0.35.}
\label{civ_pie_emissivity}
\end{figure*}

\subsection{Self-shielding} 

Our simulation assumes a uniform ionizing background which heats the gas as if it were optically thin everywhere\footnote{In practice, regions with $n_{\rm  H} > 0.1$cm$^{-3}$ (and $T < 10^{4}$K), are described with a polytropic equation of state, and are thus effectively self-shielded since their temperature is solely a function of their density.}. This introduces two potentially important limitations to our predictions of gas emissivities: (i) the ionization state of self-shielded gas is not accurately described (the neutral fraction is under-estimated), and (ii) the temperature of self-shielded gas may be over-estimated (heating from the UV background is active everywhere). \\

Self-shielding of gas to cosmological (inhomogeneous) ionizing background is an extremely complex question which can really be addressed only with simulations which jointly solve for radiative transfer and hydrodynamics. 
There has been substantial work to accomodate for that and describe {\it a posteriori} the effect of self-shielding, either in the perspective of absorbers \citep[e.g.][and references therein]{schaye2001b,schaye2001a,pontzen2008}, ``cooling'' radiation \citep[e.g.]{furlanetto2004,faucher2010}, fluorescence \citep{kollmeier2010}, or of course in the perspective of high-redshift reionization \citep[e.g.][and references therein]{aubert2010}. Such modeling is beyond the scope of the present paper. Instead, following \citet{furlanetto2004}, we chose to bracket the real solution with a series of simple assumptions : (i) a criterion to decide which gas is self-shielding (depending on the emission line of interest), and (ii) crude estimates of how self-shielding gas should emit. 

\vskip 0.3cm \noindent {\bf Self-shielding thresholds} \\ \noindent
Because we are interested in emission of IGM or CGM gas only, we first discard the gas with $n_{\rm H} > n_{\rm ISM} = 0.1$ atoms/cm$^3$. In our simulations, this gas is star-forming ISM. In the approach by \citet{furlanetto2004} e.g. it is decided that gas with $n_{\rm H} > n_{\rm SS}$ and $T < T_{\rm SS}$ is self-shielding. Here, $T_{\rm SS}$ is a temperature above which the emitting species (e.g. H{\sc i}) is mostly ionized by collisions (e.g. at $T_{\rm SS} \sim 10^{4.5}$, H is mostly in the form of H{\sc ii}, so that the gas becomes transparent to H{\sc i}-ionizing photons). Above $T_{\rm SS}$, the emitting species cannot self-shield, and photo-ionization becomes dominant. $n_{\rm SS}$ is some typical density at which self-shielding should be effective. In reality, self-shielding is a column density effect, and using a cut on density only is bound to be a poor proxy.\\ 
Here, we have in addition chosen another, more physically motivated approach. We assume that only gas at a sufficiently high pressure self-shields, and take the thermal pressure (P/k $\sim$ nT) as a proxy for the total pressure that guarantees  hydrostatic equilibrium (including e.g. magnetic pressure, cosmic ray pressure etc.). Empirically, it is found that a threshold of P/k = 155 cm$^{−3}${} K results in an HI density of $\rho = 6.1 \times 10^{7} h M_{\odot} Mpc^{−3}$ (at z=0, \citet{popping2009}), a value that is similar to the value observed in Zwaan et al. (2003) for their self-shielding thresholds. Furthermore, only gas whose radiative recombination time ($\tau _{rec} \sim T^{x} \times n_{e}$) is shorter than the sound crossing time ($\tau _{s} = R/C_{s}$, with the sound speed $C_{s} \sim T^{0.5}$ and the relevant length scale R) can recombine, and hence self-shield. A priori, it is not clear which length scale should be chosen, here we have assumed two values (R=1.0 and R=0.1 kpc). The densities and temperatures delineated by these boundaries are within the ranges given in the literature (e.g. \citet{zwaan2003, weinberg1997, wolfire2003}{} and \citet{pelupessy2005}). For the metal line transitions similar arguments hold. Further details of this method to deal with the self-shielding limits are listed e.g. in \citet{popping2009}.

These thresholds are summarised in Table \ref{tab:models}, and graphically represented by the dashed and solid lines in Figs. 1-6. Note that while we have analysed the simulation output subjecting it to a larger number of such cuts (which themselves are not part of the modelling, yet simply a-posteriori imposed cuts on certain grid cells, whose emission we either modify or ignore completely), we will focus in the following on the two most extreme cases for bracketing (subsequently called CUT 7 and CUT 8), in order to simplify the presentation.

\vskip 0.3cm \noindent {\bf Emissivity of Self-shielding gas} \\ \noindent
We use two simple scenarios to compute the emissivity of the self-shielding gas. Either we assume this gas to be in collisional ionization equilibrium (CIE) at the temperature and density provided by the simulation and use CIE tables to predict its emissivity. Or we argue that had the simulation taken self-shielding into account, the gas would be at much lower temperatures than predicted, and would simply not emit even in CIE. In that case, we set the emissivity to zero. 

\subsection{Other emission processes: photon pumping for Lya radiation}\label{sec:pumping}

There is deliberately no light emission associated with the formation of stars in this simulation. However, the UV background used here \citep{haardt2001}{}, does include on top of the QSO emission an integrated contribution from stars, but only above the Lyman limit (ie. zeroed escape fraction below that limit). Due to the large cross-section for absorption of hydrogen gas at Lyman line wavelengths, continuum photons from the non-ionizing background can be scattered when they are redshifted to the Lyman resonance lines. This process, often coined 'photon pumping', has not been accounted for by our simulation.\\

With an isotropic background, as assumed here, this pumping does not have a net effect for Ly$\alpha$ emission, as it simply consists of a redirection photons in a medium that is permeated by an already isotropic radiation field - the only true additional contribution may arise from the conversion of higher order Lyman line photons into Ly$\alpha${} photons. \citet{furlanetto2004} estimate this contribution to remain below the level of a few percent. Hence, our calculations of the gas emissivity explicitly avoid the effects of photon pumping by turning this feature off in CLOUDY, as suggested by \citet{furlanetto2004}. Note, however, that the situation changes {\it locally}, if the UV radiation field is anisotropic and potentially of different spectral shape than the general background - as it can be imagined near UV bright galaxies. Properly modelling the photon pumping contribution (or actually tracing the continuum photons responsible for it on their paths through the simulated cubes) is beyond the scope of this paper, but we have performed  calculations with simplified assumptions to judge the importance of it (cf. 6.2). From these, we conclude that indeed enhancements of the flux over the emissivities calculated here of an order of magnitude are possible in the proximity of UV bright sources.       

\begin{table*}
 \caption{Different models for the emissivity of the gas in the light cones. Generally speaking, we divide the parameter space in loci that exhibit either strictly no emission, or an emissivity is calculated by applying CLOUDY with either photo-ionisation equilibrium (PIE) or collisonal-ionisation equilibrium (CIE). The number indicated for each model corresponds to the one used in the different plots. Models 2 and 3 follow closely the procedure and values suggested by \citet{furlanetto2004}, while models 4 and 5 allow for a ten times higher limiting density threshold for CIE or PIE. Models 6-9 involve the pressure and timescales for recombination and sound crossing, as described in the text and in \citet{popping2009}, for different size scales (R=0.1 and R=1.0 kpc). The units for the density are [n]=cm$^{-3}$, for the temperature [T]=K, and [P/k]=cm$^{-3}$K. As figures 1-6 show, the cuts labeled 7 and 8 are the most severe and lenient, respectively, for the line emission, and hence we use them to bracket the intermediate cases.}
 \label{tab:models}
 
 \begin{tabular}{@{}lcccc}
          & Cut & Zero emissivity & PIE emissivity & CIE emissivity \\
   \hline
   SF cut & 1 & $n_{\rm H} \geq 0.1 $ & $n_{\rm H} < 0.1 $ & - \\
   +CIE & 2 & $n_{\rm H} \geq 0.1$ & log T$>$4.5 OR                                  & log T$<4.5$ AND \\
                       &                     & $n_{\rm H} < 5.1\times 10^{-3} (\rmn{Lya})${}   $^{1}$  & $n_{\rm H} \geq  5.1\times 10^{-3} (\rmn{Lya}) $ \\
   
   +cut & 3 & $n_{\rm H} \geq 0.1$ OR & log T$>$4.5 OR  & - \\
  &  & (log T$<4.5$ AND $n_{\rm H} \geq 5.1\times 10^{-3} (\rmn{Lya}) $) & $n_{\rm H} < 5.1\times 10^{-3} (\rmn{Lya}) $ & -  \\
   +CIE x10  & 4 & $n_{\rm H} \geq 0.1$ & log T$>$4.5  OR                & log T$<4.5$ AND \\
                &                     & $n_{\rm H} < 5.1\times 10^{-2} (\rmn{Lya}) ${}  $^{2}$  & $n_{\rm H} \geq   5.1\times 10^{-2} (\rmn{Lya}) $ \\

   +cut x10  & 5 & $n_{\rm H} \geq 0.1$ OR & log T$>$4.5 OR  & - \\
   & & (log T$<4.5$ AND $n_{\rm H} \geq 5.1\times 10^{-2} (\rmn{Lya}) $) &  $n_{\rm H} < 5.1\times 10^{-2} (\rmn{Lya}) $ & - \\
   \hline
   R=1 kpc + CIE & 6 & $n_{\rm H} \geq 0.1 $ & P/k $<$258 OR $\tau _{rec} > \tau _{s}${}  $^{3}$  & P/k $>$258 AND $\tau _{rec} < \tau _{s}$ \\
   R=1 kpc + cut & 7 & P/k $>$258 AND $\tau _{rec} < \tau _{s}$  & P/k $<$258 OR $\tau _{rec} > \tau _{s}$ & - \\
   R=0.1 kpc + CIE & 8 & $n_{\rm H} \geq 0.1 $ & P/k $<$258 OR $\tau _{rec} > \tau _{s}$  & P/k $>$258 AND $\tau _{rec} < \tau _{s}$ \\
   R=0.1 kpc +cut& 9 & P/k $>$258 AND $\tau _{rec} < \tau _{s}$  & P/k $<$258 OR $\tau _{rec} > \tau _{s}$ & - \\\hline
\end{tabular}

\medskip
$^{1}${} The values change with different redshifts, hence : $n_{\rm H} \geq 7.1\times 10^{-3} (OVI)${} and $n_{\rm H} \geq 2.7\times 10^{-3} (CIV)$.\\
$^{2}${} $n_{\rm H} \geq 7.1\times 10^{-2} (OVI)${} and $n_{\rm H} \geq 2.7\times 10^{-2} (CIV)$.\\
$^{3}${} For OVI at z$\sim$1.0, we apply P/k(z) = 356, and for CIV at z$\sim$0.35  P/k = 135 as dividing line. \\

\end{table*}

\subsection{Simulated Observations} \label{sec:observed_cubes}

We construct mock data cubes from our simulations in two steps: first, we build a {\it pre-observation} light-cone at the full AMR resolution, and second, we convolve this light-cone with instrumental effects. 

\vskip 0.3cm \noindent {\bf Pre-observation light-cones} \\ \noindent
The typical geometry of the light-cones we wish to produce is that of a pencil beam, i.e. a cone with a radial extent much longer than the size of our simulation (typically $\geq 5$ times) and an angular extent probing scales less than a tenth the size of our simulation at most.  Another point in consideration is that we are interested in emission from diffuse intergalactic gas, which can produce extended signal on scales up to a few tens of Mpc. It is thus important to preserve the continuity of the gas density field in the light-cones. We use the MOMAF software \citep{blaizot2005} to cut a light-cone out of our simulation. MOMAF takes care of replicating our (periodic) simulation at will and to cut light-cones in any direction relative to the axis of the box. In order to preserve continuity, we chose not to use the random tiling technique and to build light-cones out of one simulation output only. This avoids both redshift and replication discontinuities \citep[see][for details]{blaizot2005}. We then chose lines of sight by carefully selecting angles for light-rays in order to minimize replication effects. 

In practice, we build three different types of data cubes (see below, one for each emission line of interest).\footnote{Here, and in the following, we use the term 'data cube' for a specific output of the simulation, where we have taken a part of the whole simulated volume and processed it in different ways. For some of the latter analyses, e.g. we have created regularly-gridded data cubes, whereas other aspects of the analysis may rely upon the original AMR gridding. As such, the terms 'cone' and 'cube' are for practical reasons equivalent in their meaning.} For this, we use the snapshots, i.e outputs at $z \sim 0.35$ (resp. 0.75, 1.04) for C{\sc iv} (resp. H{\sc i}, O{\sc vi}). We have checked that our conclusions are not changed if we use slightly different snapshots, or if we use them all and break the continuity of the density field. 

In the outputs, each gas element (cell) has a luminosity distance, an apparent redshift (taking into account the gas' peculiar velocity), and angular coordinates.

\vskip 0.3cm \noindent {\bf Observed data cubes} \\ \noindent
As an instructive example, we chose to mimick the instrumental characteristics of FIREBall \citep{tuttle2010, milliard2010}. We will discuss alternative choices of angular or spectral resolutions in Sec. 5, but note here that current or near-future UV missions are of quite similar layout. While focusing on the FIREBall spectral coverage forces certain specific redshift choices for the three lines in question, it does also allow us to check on differences regarding the cosmic evolution over that redshift span ($0.35 < z_{em} < 1.1$).    

In order to produce observed data cubes from of our pre-observation catalogs, we proceed in 3 steps. First, at the full AMR resolution, we compute the luminosity of each cell according to the various models described in Sec. \ref{sec:gas_emission}. Second, we derive each cell's flux contribution to a regular grid (in $\alpha, \delta$, and z) with better resolution than the typical angular and spectral point-spread function. Third, we then convolve these in the spectral as well as the spatial dimensions to arrive at the  desired final resolution (in angular direction we apply a sharp disk of an 8 arcsec diameter, and in the spectral direction we convolve with a Gaussian of FWHM = 2.25 pixels $\sim$ 0.56 \AA), and finally save the convolutions on a regularly gridded data-cube. 

We have generated 300 of such fields of view with an approximate 3x3 arcmin$^2$ FOV (similar to FIREBall's FOV) in order to assess the effects of cosmic variance (e.g. for the construction of the PDFs). Furthermore, for instructional purposes, we have specifically chosen 3 sightlines (for each transition and redshift) with a larger FOV : one towards a field containing a known very massive Dark Matter structure to represent a best case-scenario for 'bright' sources, one which contains an example for a bright, large filamentary structure, and another one as the 'best' compromise between those two criteria. These fields cover a field-of-view (FOV) of fixed size (900 $\times$900 arcsec$^{2}$), and fixed spectral coverage (270 \AA{}), thus creating cubes of different volumes depending on the specific redshift. The selection of these sightlines, in fact, used PDFs (see below) from fields of smaller FOV, in order to determine their brightness distributions. \\

Figures 1 to 3 show the phase space diagrams of the simulation, each cell weighted by its luminosity in the  Ly$\alpha$, OVI or CIV transitions. Note how dominant the high density areas (log n$_{H} \geq -2$) are in these, as emissivity (denoted by the solid, curved lines) scales with $n_{H}^{2}$. It is immediately obvious that the treatment of  self-shielding  is very important for the Ly$\alpha${} estimates, while it plays less of a role for the metals, which emit most of their light from gas at temperatures $\geq 10^5${} K.


\section{General Features of the Line Emission}\label{sec:general_features}

\begin{figure*}
\includegraphics[width=168mm]{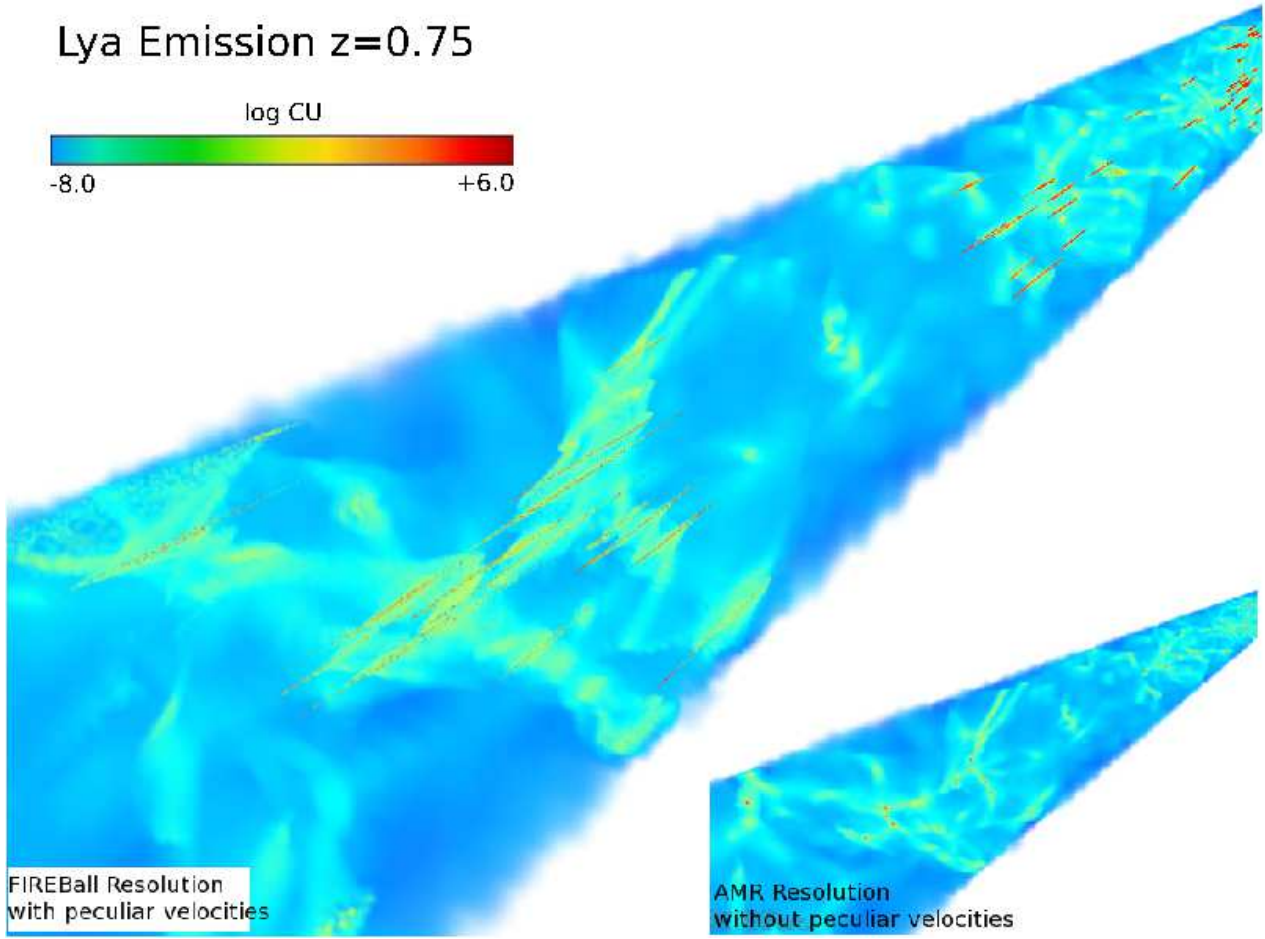}\caption{Examples of the simulation output for Ly$\alpha${} line emission at z=0.75. The lower right inset shows a part of an AMR lightcone with an opening angle of 15$\times$15 arcmin$^2$ in the x-y direction and a length in the z direction of $\sim$100 \AA\, corresponding to a structure in (physical units) of 6.6 $\times$ 6.6 $\times$ 300 Mpc. The web-like structure with long, elongated filamentary bridges connecting the bright nodes (red labelling) is clearly distinguishable. The top panel shows the same volume, but now both degraded to a resolution resembling currently available instruments (i.e. about 8 arcseconds in spatial, and 0.5 \AA\ in wavelength direction) and with peculiar velocities added to each cell. Note how these tend to elongate the bright sources in z-directions, while leaving the filamentary structures almost untouched. Also note how the bright sources are clearly clustered, and not randomly distributed.}\label{map}

\end{figure*}

Even a casual visual inspection of the line emission inside the simulated cubes (cf. Fig. \ref{map}) reveals a clear bimodality in the types of emitting regions. There are on the one hand bright sources, spatially confined to small areas, but extended in wavelength due to the gas being spread out in velocity space. In the vicinity of these, and connecting such regions, there are filamentary structures, which are much fainter, more extended spatially and affected less by the velocity dispersion of the gas. We will discuss the specific properties of these two different classes of emitters in the next two sections, but focus here first on general implications.\\

\begin{figure*}
\includegraphics[width=168mm]{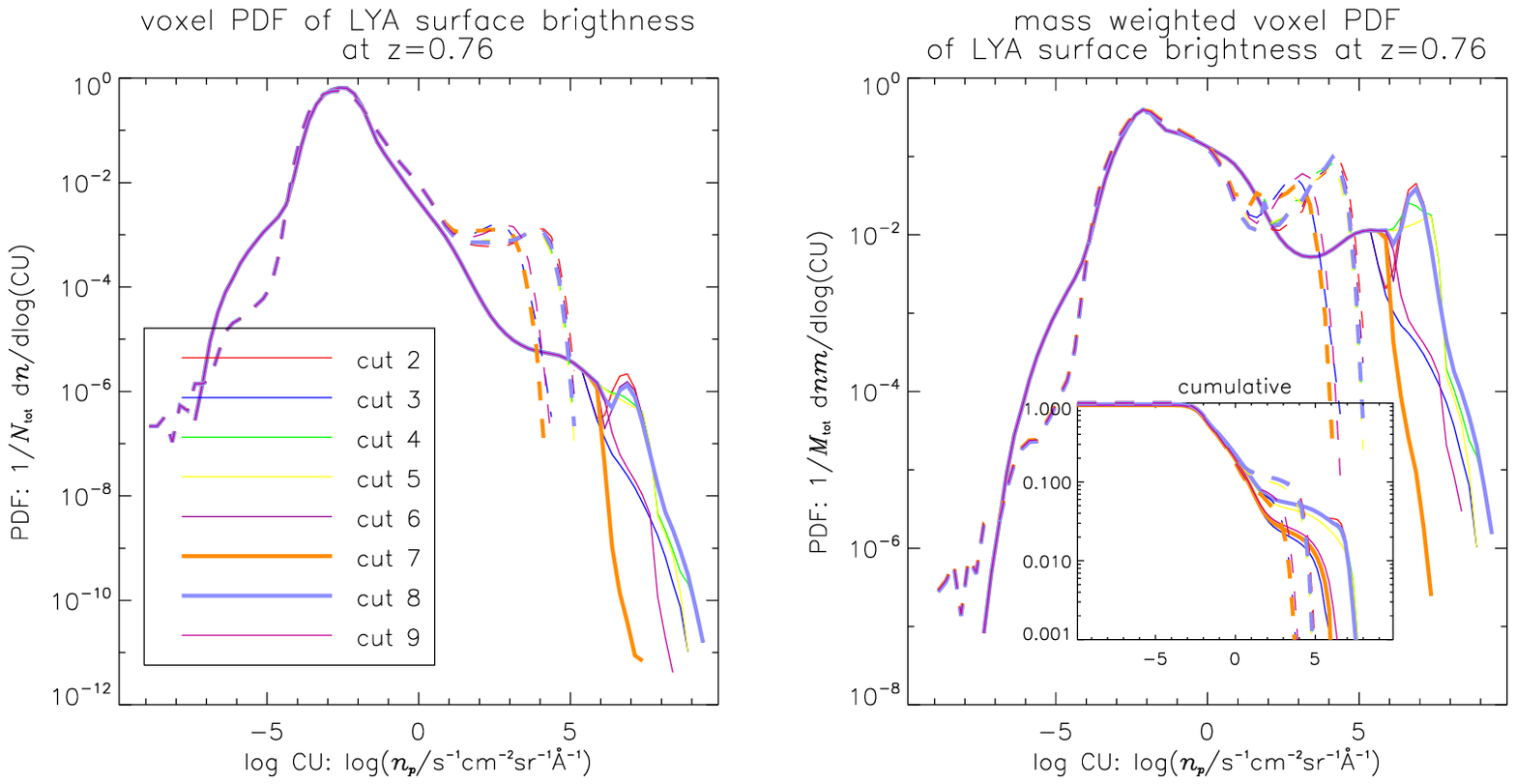}\caption{The probability distribution function (PDF) of the voxels' line emission in Lya for z$\sim$0.75, weighed by volume({\bf left panel}; i.e. we simply count all the cells' volume per certain brightness bin and divide by the total survey volume in order to derive the fraction per bin). Note that this is hence a 'true' pdf in the sense that we are using the complete 3D cube, and do not collapse along one dimension (in contrast to e.g. the PDFs by \citet{furlanetto2004} or \citet{bertone2010b}, whose results are based on the pixels of an image). The sudden  change in slope when reaching the plateau around a few CUs from the low surface brightness side is indicative of reaching a different population, which can - as detailed in the text and Figs. \ref{example_lya_075} and \ref{example_lya_1165}  - be equated with bright compact sources.  The effect of resampling from the AMR refinement level (solid lines) to the observationally dictated resolution (dotted lines) is a  dramatic loss of the highest surface brightness objects. Note also that the choice for different treatment of self-shielding results in factors of 10-100 for the brightest spots.  Bright sources are rare, but carry almost all of the light : 10$^{-4}$ of the volume after resampling. The most important aspect regarding the resampling, however, is that the very brightest spots still remain within the range of capabilities of an instrument with current technology, as discussed in section \ref{sec:observing_strategies}. The {\bf right panel} shows the PDF weighed by mass, the differential mass distribution being the large plot, and the inset representing the cumulative distribution. It is clear that being able to trace a sizeable fraction of the baryonic mass ($\sim 10\%$), an instrument needs to go down to a surface brightness limit of a few CU.}\label{lya_pdf_volume}
\end{figure*}

\begin{figure*}
\includegraphics[width=168mm]{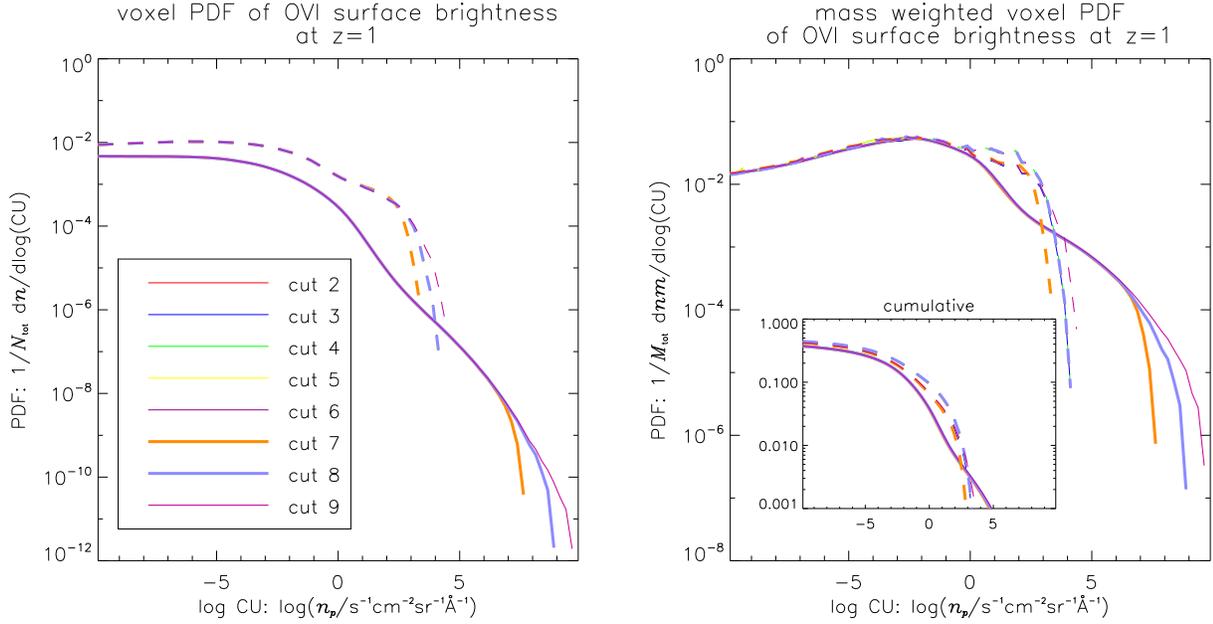}\caption{Same plot as Figure \ref{lya_pdf_volume}, but this time for OVI emission at z=1.1. In contrast to the Ly$\alpha${} radiation, here most of the cube's volume is filled by gas in a regime (low temperatures, low densities and extremely low, if not zero, metallicity) that does emit very few, if any photons, hence the integral over the surface brightness range depicted here does not add up to unity. Note how much less important the cuts for self-shielding are regarding the distribution of the brightest voxels in the metal line, indicating that the main contribution to the flux in OVI (and CIV, see Figure \ref{civ_pdf_volume}) is emitted by gas with temperatures higher than the limits imposed by our self-shielding treatment. Note also how the some of the curves are exactly overlapping, as the self-shielding cuts do not affect any areas contributing to the surface brightness levels shown here.}\label{ovi_pdf_volume}
\end{figure*}

\begin{figure*}
\includegraphics[width=168mm]{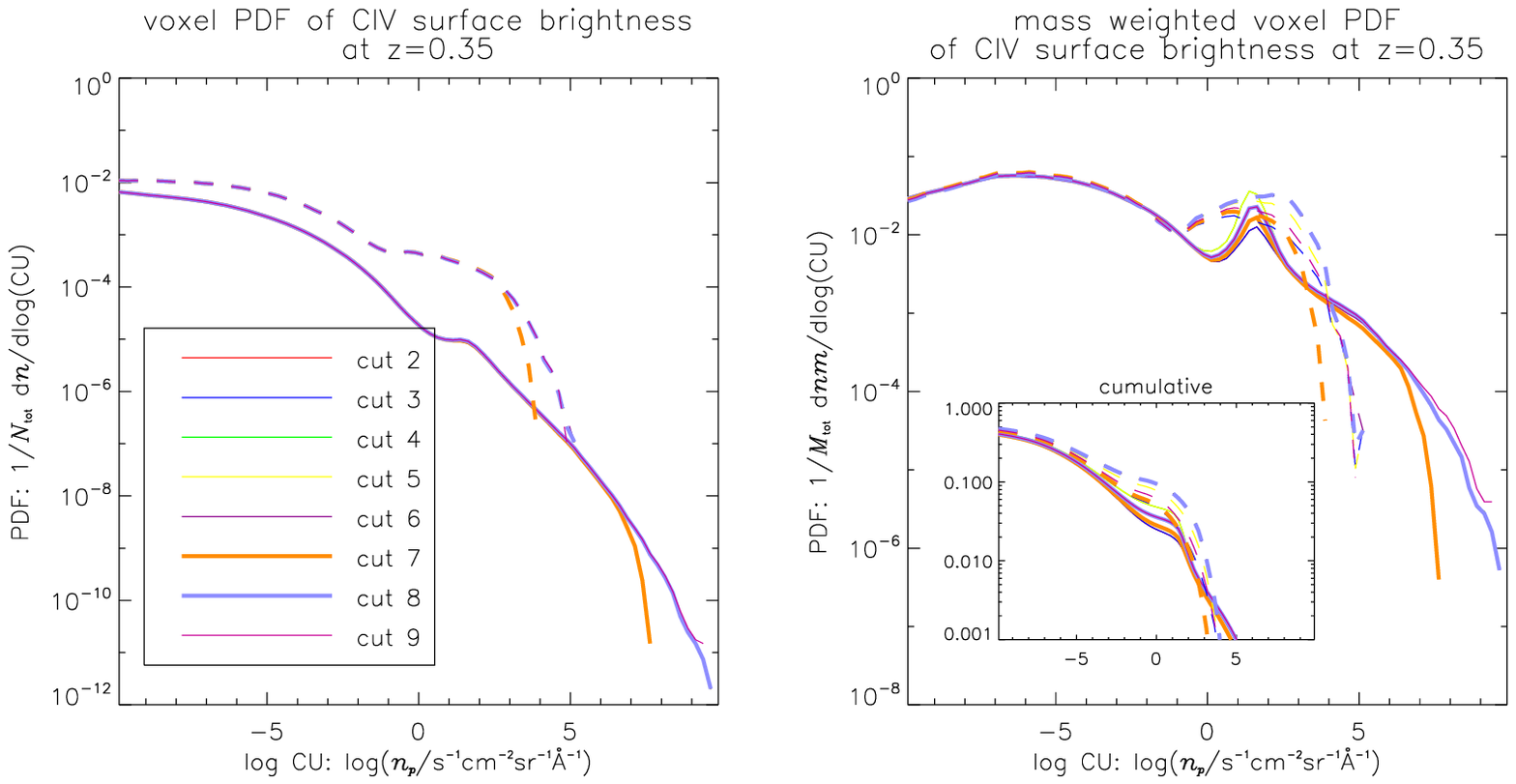}\caption{Same as Figure \ref{lya_pdf_volume}, but for CIV at z=0.35.}\label{civ_pdf_volume}
\end{figure*}

Figures \ref{lya_pdf_volume}, \ref{ovi_pdf_volume}, and \ref{civ_pdf_volume} show the probability distribution functions (PDF) of the voxels' emission in Lya, OVI and CIV, weighted by volume. The solid lines represent the PDFs based upon the full AMR resolution, while the dashed lines are PDFs calculated after resampling of the data cubes with a grid of 3 arcseconds spatially and 0.25 \AA{} spectrally, subdividing the simulated cube smoothed with the resolution of 8 $\times$ 8 arcsec$^2$ and 0.56 \AA\ as described earlier (resembling the FIREBall instrument specifications). Note that unlike e.g. the PDFs presented by \citet[]{bertone2010b} and \citet[]{furlanetto2004} these are PDFs based upon the full 3D voxel distribution, i.e. we do not collapse along one axis and derive the distribution on the resulting 2D images of such slices. Hence, a comparison of the curves is not straightforward. The various cuts to deal with the self-shielding gas are indicated by the different colours. Obviously, the two extreme distributions are achieved when the cuts are most lenient or most severe in excluding gas located in the parameter space that exhibits the highest emissivity. The former happens when we choose to apply the CIE case for R=0.1 kpc (denoted as example 8 in Table 1), the latter is achieved by neglecting all emission for R=1.0 kpc (example 7). In the following, and in various graphs, we will refer to these two extreme scenarios as CUT7 and CUT8, respectively. \\

A decrease in resolution has a dramatic effect on the brightest areas, but leaves the distribution unchanged for surface brightnesses less than $\sim 1${} CU\footnote{1 CU = 1 photon s$^{-1}$cm${-2}$sr$^{-1}$\AA${-1}$}. The upward shift of the plateau ranging from about $2.0 \leq$ log CU $\leq 7.0$ at the AMR level to a similar plateau with a lower cutoff high surface brightness, but a higher volume fraction at the lower resolution clearly indicates that the brightest pixels are strongly 'clustered', i.e. form collectively the bright, spatially compact sources. These bright regions are rare : even for the resampled cubes they do not cover more than 0.1\%{} of the volume, but carry almost all of the light emitted in all three lines. The various cuts dealing with the self-shielded gas allow us to bracket the expected emission. Note that in the case of Lya the cut-off values for the brightest sources change by a factor of 100, a direct reflection of the fact that this is the location of the parameter space where the emissivity for Lya peaks (cf. Fig. \ref{lya_pie_emissivity}). In contrast, as expected, the exact treatment of the self-shielded gas has very little effect on the bright end of the surface brightness distributions for the OVI and CIV line emission.\\


\begin{figure*}
\includegraphics[width=168mm]{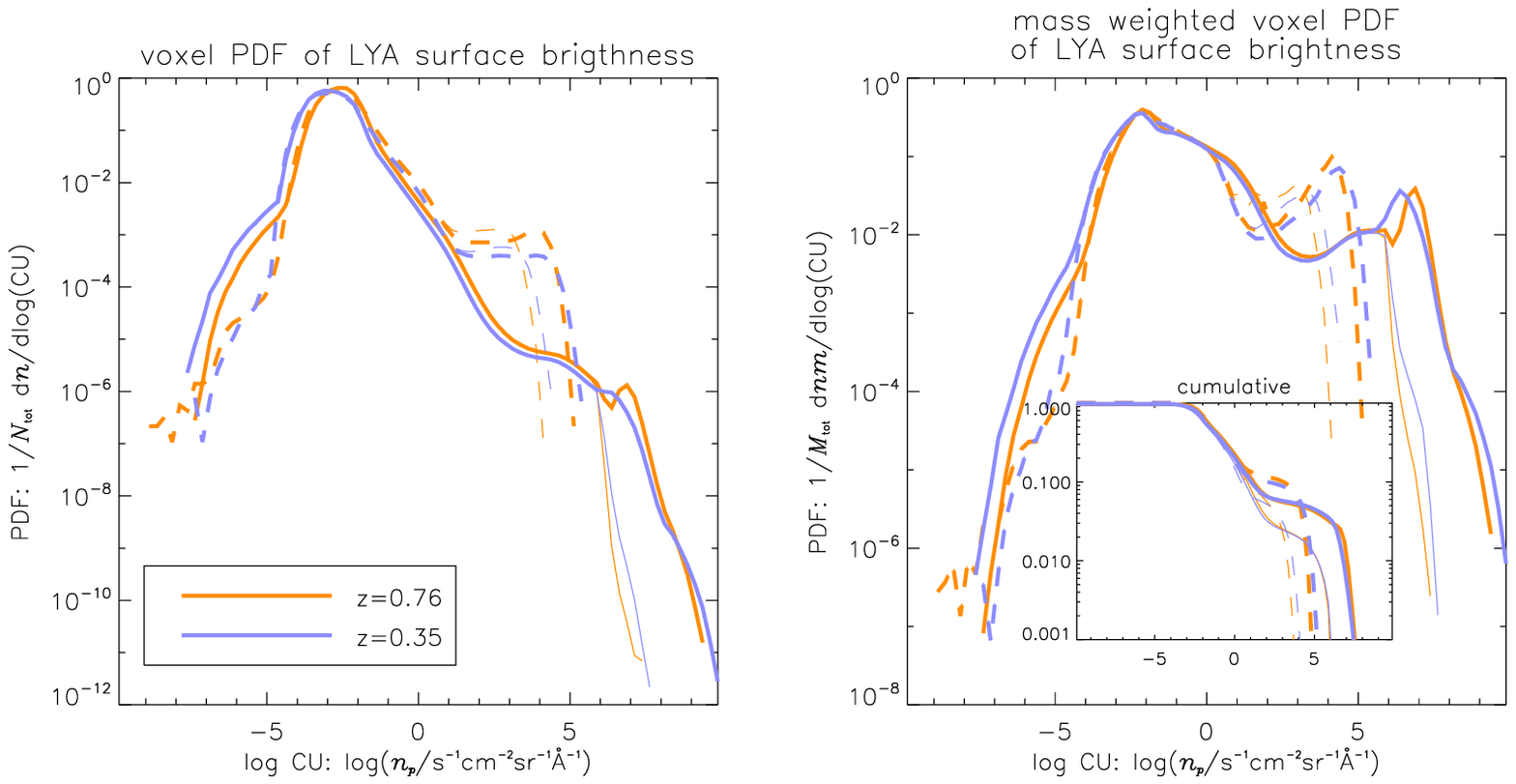}\caption{Similar to  Figure \ref{lya_pdf_volume}, but now emphasising the redshift evolution for the Ly$\alpha${} emission from z=0.75 to z=0.35, and employing the two most extreme treatments for the self-shielding gas. The peak of the distribution shifts to lower surface brightnesses with decreasing redshift, mostly due to a weakening UV-background and dilution of the gas because of cosmic expansion. The bright, isolated areas show a slight increase in their brightness, but not as much as expected from a simple (1+z)$^{4}$\ scaling, rendering the evolution a more complicated function with redshift, density, temperature, and environment contributing to the complexity.}\label{pdf_volume_diff_z}
\end{figure*}

Figure \ref{pdf_volume_diff_z}{} shows two PDFs for Ly$\alpha${} at the redshifts of z=0.75 and z=0.35. Note that there is by no means a simple (1+z)$^{4}$ evolution at all, as may be expected from sheer
cosmological expansion effects (meaning that the lower redshift values should exceed the higher ones by a factor of about 3), but a more complex evolution. The faint emission peak e.g. becomes even less bright, mostly due to the change in the ionising radiation field weakening towards lower z. The brighter regions beyond the knee in the distribution, however, do become brighter at the lower redshift. Hence, the redshift evolution of the surface brightness distribution is a complicated function, whereby geometric cosmic expansion and true evolutionary source effects combine in a non-trivial fashion. In any case, the evolution expected from redshift changes surveyed here is probably small compared to the uncertainties introduced by the self-shielding treatments. 
  
\section{Bright, compact Sources/CGM}\label{sec:compact_sources}
\subsection{Introduction to the source finder}
The simulations with the reconstructed distribution of Ly$\alpha$, CIV and OVI contain bright objects with extended environments connected by filaments spanning up to the whole  spatial scale of the simulation output. Features in the simulated outputs only represent circum- and inter galactic gas, as all the star forming gas and gas in the galaxies has been removed using thresholds as described in 2.2. To extract objects from the simulated cubes, we have used the source finding algorithm {\it Duchamp} \citep{whiting2008}, an algorithm that has specifically been developed to detect objects in 3D radio observations that typically have dimensions of RA, DEC and velocity. The software is however very flexible and can be used for any kind of three dimensional dataset.\footnote{The algorithm is decribed and can be obtained here : http://www.atnf.csiro.au/people/Matthew.Whiting/Duchamp/} \\

The simulation outputs tailored to the spectral and angular resolution specifications of FIREBall (see earlier) were used as input for {\it Duchamp}. Within {\it Duchamp}, a parameter input file constrains the properties of a possible detection. For all objects a  requirement has been set, that  detections consist of at least two adjacent voxels in the spectral direction. Because of the oversampling in the spectral direction by about a factor two, this is the minimum size an object can have.\\

Within {\it Duchamp} objects were sought using two different thresholds. The first threshold represents the minimum peak flux a feature should have to be accepted as a compact object. In the reconstructed Ly-$\alpha$ cubes compact objects can be distinguished by their large peak flux. The brightest regions in the filaments have typical values of the order of $\sim 10$ CU, therefore the first threshold for Ly-$\alpha$ is chosen at 100 CU.  Once the location of bright and compact features has been determined, the objects are grown in size by adding adjacent voxels until a second (lower) threshold has been reached. By using two different thresholds we can isolate compact objects down to a relatively low flux threshold which is below the peak flux of extended filamentary structures. For Ly-$\alpha$ the second threshold is chosen at 10 CU. Although detections could be extended to lower flux thresholds, this would increase dramatically their size and include parts of filaments or possibly even merge them, without adding significant amounts of flux. \\ 

For each transition the threshold fluxes were determined empirically by carefully inspecting the cubes, the exact values are given in table ~\ref{thresholds}. The sharp edges of the source regions make the {\it number} of the detected sources largely insensitive to the threshold values, as it is possible to choose a value for the second threshold that allows for picking up a large fraction of the total light inside the simulated volume while ensuring that there is little source merging. The same flux thresholds have been used for all self-shielding cuts, since the selected cuts only affect the densest regions, and even the most severe such cut does not prevent detection with the parameters chosen here.\\

\begin{table}\caption{Flux thresholds used within Duchamp to find detections in  the simulated cubes of Ly-$\alpha$, CIV and OVI. }\label{thresholds}
\begin{tabular}{lcc}
\hline
\hline
Element & threshold 1 [CU] & threshold 2 [CU] \\
\hline
Ly-$\alpha$ & 100 & 10 \\
OVI & 1 & 0.5 \\
CIV & 0.5 & 0.05 \\
\hline
\hline
\label{thresholds}
\end{tabular}
\end{table}

\subsection{Source properties : Space density, luminosities, sizes, shapes, spectral information}

 The number, and hence space density (not the flux) of those sources in Ly$\alpha${} is very robust and almost independent\footnote{For the harshest cut criteria, a select few (in the extreme case 4 out of $>1000$) Lya objects fall below the minimum detection threshold that could have been picked up for the more lenient cuts.}  of the self-shielding cut criteria {\bf and}{} the specific parameters we use for the detection algorithm, and thus a strong prediction of the simulation. We find a number density for the Lyman $\alpha${} bright sources of $\eta (Lya, z=0.75) = 38 \times 10^{-3}$ (Mpc/h)$^{-3}$. Interestingly, the densities for bright sources in CIV and OVI are not very different : $\eta (CIV, z=0.37) =  24.8 \times 10^{-3}$ (Mpc/h)$^{-3}$, and $\eta (OVI, z=1.1) = 17.3 \times 10^{-3}$ (Mpc/h)$^{-3}$. In these cases, however, the detection algorithms' parameters do lead to a fraction of the faintest sources being dropped, i.e. it is not possible to find good combinations of the 2-fold threshold approach that isolates the point sources without introducing overlaps and/or losses. As those sources are too faint to be detected in any realistic scenario (see paragraph 6.2), we have not tried to fine-tune the detection algorithm in order to better isolate these objects. In any case, of much more importance, even for Ly$\alpha${} compact sources, are the restrictions imposed onto us by the simulation's mass and spatial resolution. We have performed tests regarding conversion with earlier, lower resolution simulations, and found that we are 'complete' to a luminosity of log L(Ly$\alpha$)$\geq$ 41.5 (at z=0.75; resp. 41.2 for z=0.35), log L(OVI, z=1.1)$\geq$ 40.5, and log L(CIV, z=0.35)$\geq$40.0 (see below for calculation of the source luminosities). That is, increased resolution may result in producing a higher fraction of areas that had fallen underneath the threshold criteria used here in the lower resolution simulations, due to the strong non-linearity of the brightness as a function of density and temperature. Thus, for all transitions, the numbers quoted here are strictly lower limits. Furthermore, we remind the reader that we have extracted these sources in three different cubes for each transition that were not randomly chosen, but care was taken to select sightlines that do show large-scale structure. While this procedure resembles in some respect the target selection for a real observation, we nonetheless point out that cosmic variance is an additional factor in producing fluctuating number densities for which we have not accounted here. \\

\begin{figure*}
\includegraphics[angle=270,width=168mm]{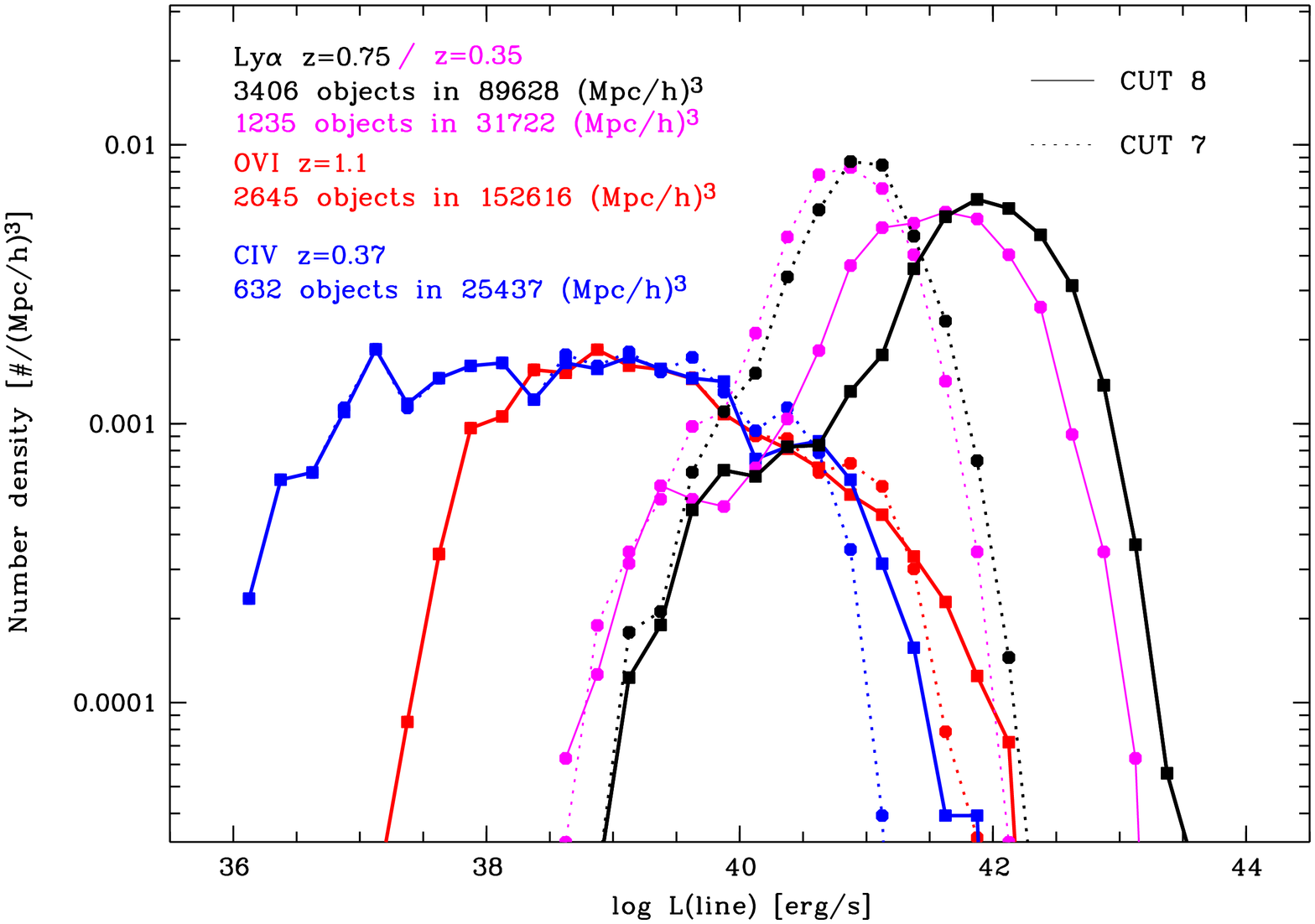}\caption{The luminosity distribution of sources extracted by Duchamp of cubes for three different transitions and redshifts, plotted differentially as the number density per (comoving) Mpc$^{3}$h$^{-3}$ in bins of log L = 0.25. The two different line styles for each transition are for the two extreme cuts for treating the self-shielding gas (CUT 8 being the most lenient, CUT 7 the most severe ; see text for details). Note that while the cuts have no effect on the total number of sources, a dramatic effect for the source luminosity in Ly$\alpha${} is apparent, whereas the distributions for the metal lines remain virtually unchanged. For Ly$\alpha$ the cuts allow to bracket the most extreme cases. Notice how broad the distributions for the metal line transitions are in comparison to the rather sharply peaked Ly$\alpha${} luminosity distributions. This may partly be explained by compounding the complication of source 'incompleteness' due to the simulation resolution with the imperfections of the source extraction for the metal lines, whereas for Ly$\alpha${} we only need to to worry about the former (for details see text).}\label{luminosity_distributions}
\end{figure*}

\begin{figure*}
\includegraphics[angle=270,width=168mm]{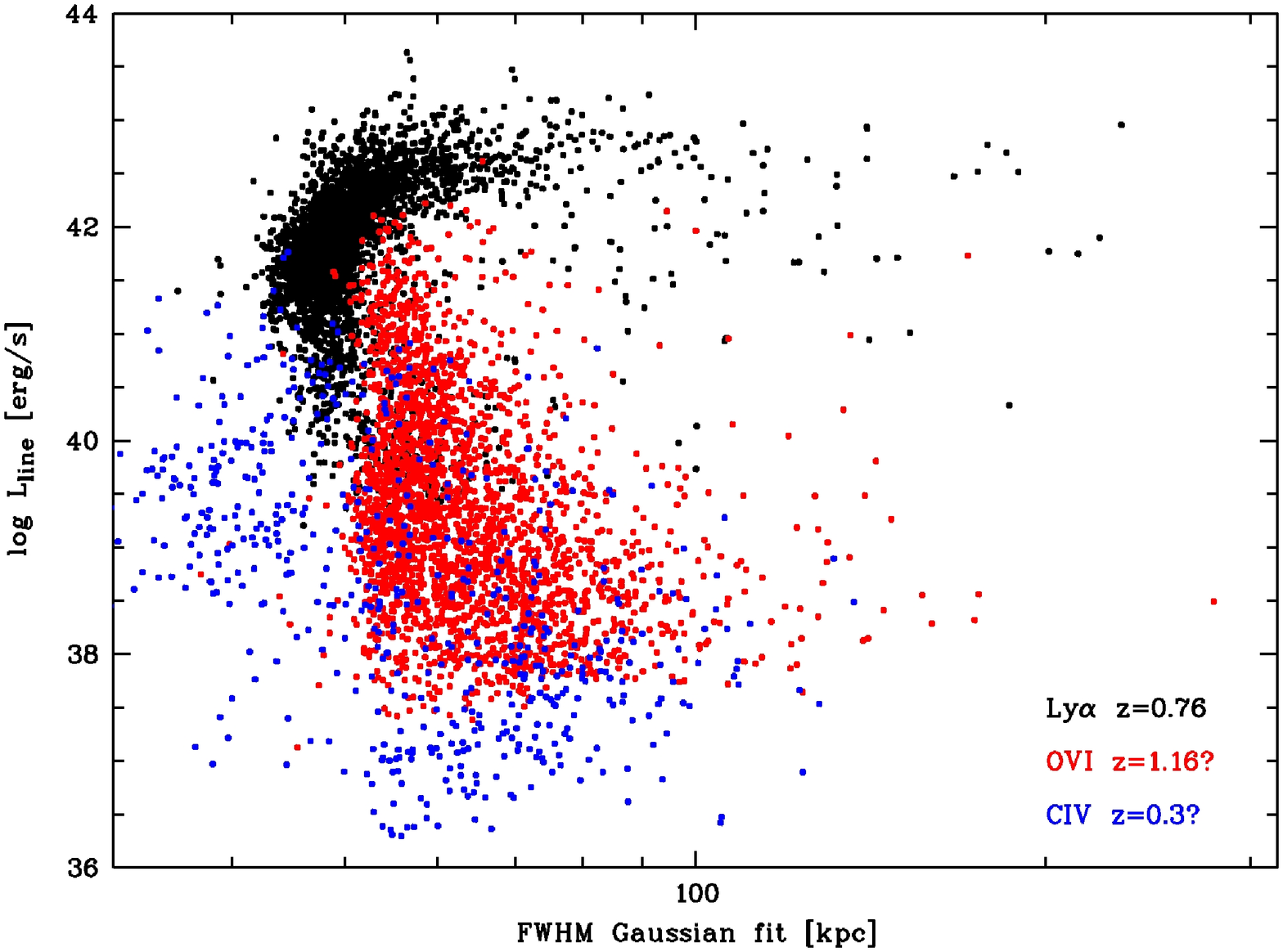}\caption{The size of the bright sources (measured as detailed in text via fitting a 2-D Gaussian to the images collapsed along z direction) vs. their luminosities. While there is a clear trend of more luminous Lya emission regions being more extended, the same does seem not to hold for OVI and CIV. The sharp cutoffs for the lower boundary of the sizes are a result of the minimum FWHM allowed being 2 pixels, translating into different sizes at each redshift. Plotted are 3400+ sources in 3 different cubes for Ly$\alpha$, and 2600+ resp. 600+ sources for OVI and CIV. All luminosities have been estimated using CUT8 (the most lenient one), and utilise the results from the Duchamp source algorithm.}\label{size_vs_luminosity}
\end{figure*}

\begin{figure*}
\includegraphics[width=168mm]{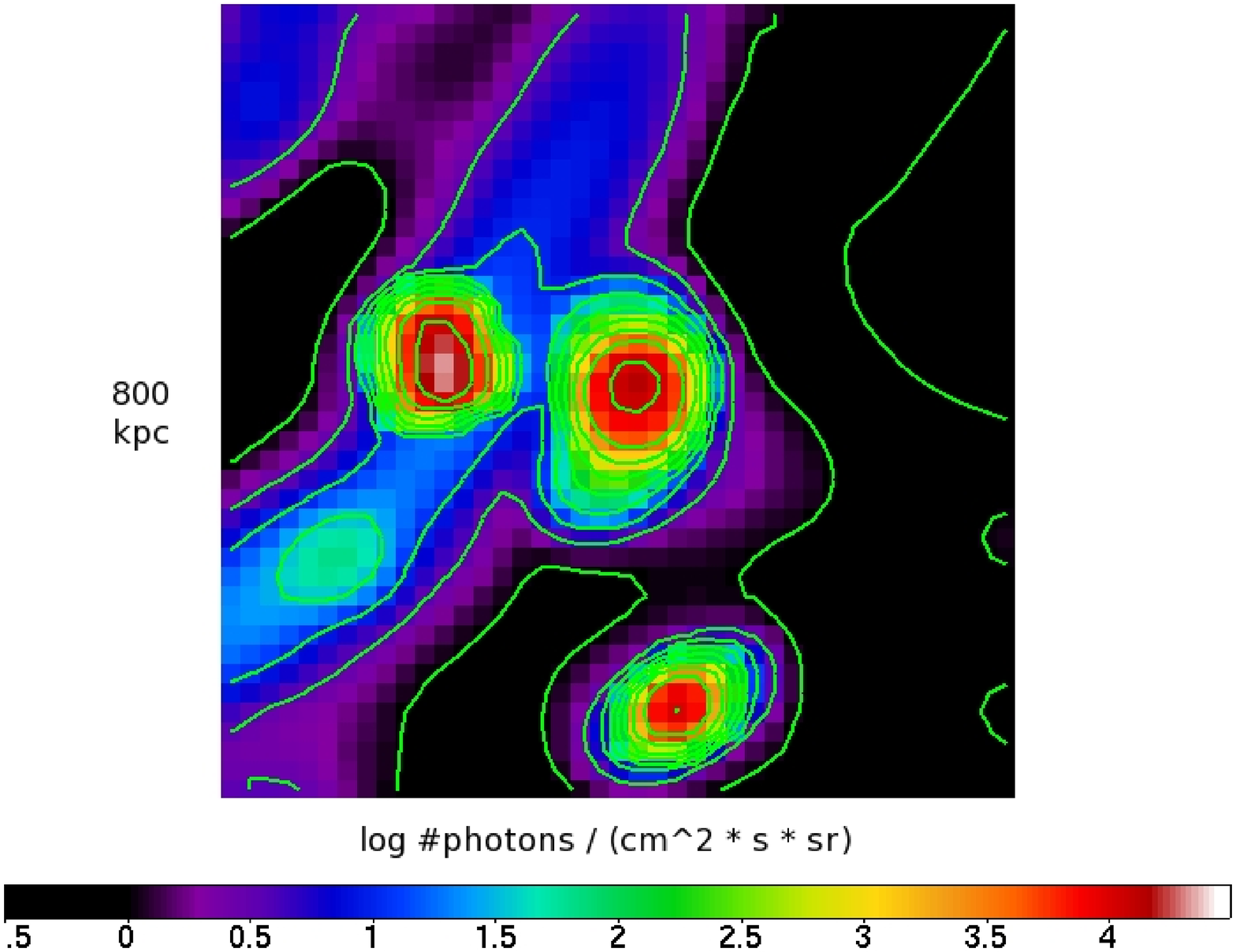}\caption{Example of a compact bright Ly$\alpha${} source at z=0.75 with a simple spatial profile (one single core ; center of image), as seen in a cube with a spatial resolution of 3 arcseconds . The image is a 40 by 40 pixel cutout, collapsed along the source's maximum extent in z direction. At this redshift, this translates to 800 (physical) kpc. The green contours delineate areas of +0.5 dex in  brightness. Note how all three bright sources in this image are clearly separated, and are up to 4 orders of magnitude brighter than their surrounding areas. The unit of brightness here is LU rather than CU as we have collapsed along the spectral axis.}\label{example_lya_075}
\end{figure*}

How big are the objects and what are their shapes ? The answers to these questions are not quite as straightforward as one may naively assume. While the spatial extent perpendicular to the sightline in our cubes immediately translates to a size measurement, the coordinates along the sightline have the additional complication that the pixels we analyse have their peculiar velocities imprinted on them. As a first step towards estimating the sizes of the sources we have hence constructed small cutouts of the cubes (centered on the coordinates found by {\it Duchamp} as central pixels) and summed them up along the full extent of the sources (given by their minimal and maximal estimate for the z coordinate in {\it Duchamp}). We then try to fit a simple 2-D Gaussian to the light profile of such a collapsed image. In the majority of the cases the fitted FWHMs of the Gaussian (in x and y direction) are of a similar size, indicating that the underlying light profiles can be reasonably well described by the average of them for a crude size measure (Figure \ref{example_lya_075}{} shows an example of such an object in Lyman $\alpha$). For some objects, however, this blunt approach fails because their spatial (and often also velocity, see below) structure is more complicated. Figure \ref{example_lya_1165}{} highlights a source that shows multiple bright cores, and even a hint for 'bridges' between them. \\

Figure \ref{luminosity_distributions}{} shows the distribution of luminosities of the distinct sources detected by Duchamp in the Lya and metal line cubes, smoothed to FIREBall resolution, and according to the two most extreme treatments for the cells containing self-shielding or star-forming gas (CUTS 7 and 8). For Lyman $\alpha${}, the luminosities of the objects picked up by {\it Duchamp} is strongly dependent on the specific treatment of the observed cubes with regards to the self-shielding and star-forming gas, as figure \ref{luminosity_distributions}{} clearly demonstrates. The various  cuts introduced earlier lead to luminosities for the same objects straddling almost a factor of 100 in brightness in the extreme cases. Hence, for the Ly$\alpha${} sources we can only bracket the luminosities assuming either the severe cuts (resulting in a median luminosity of log L [erg/s] = 40.9 ) or the optimistic cuts (median log L [erg/s] = 41.8). The estimates for the OVI and CIV emission, on the other hand, are much more robust, because - as we have seen earlier - the cuts we introduce are not affecting pixels of the highest emissivity in these two transitions (median log L[erg/s] = 39.2 and 38.5 for OVI and CIV, respectively). Notice also, that the distributions for the bright objects in the metal lines are much broader than for Lyman $\alpha${}. This is not, as we will see in the next paragraph, due to a broader distribution in sizes of these objects, but a result of a larger spread in the peak brightness of the most dominant pixels. Overall, the objects observable in OVI and CIV are in general at least 1.5 orders of magnitude less luminous than the HI sources, but are spread fairly evenly over almost 3 decades. Keep in mind, however, that for an assessment of their observability the much lower redshift for CIV and hence luminosity distance does help (yielding a factor of $\sim$7 in surface brightness).\\

 In Figure \ref{size_vs_luminosity} we show the relation between the luminosities and the FWHMs (averaging the x and y components) of the compact objects - note that we flagged out the obvious failures of the simple fitting. While the majority of the sources lie between about 50 and 100 kpc (proper), there are a few exceptions that can reach up to the enormous size of 300 kpc. The sharp low cutoff boundary (most easily visible for HI and OVI) represents the minimum FWHM of 2 pixels, i.e. is a result of the image resolution.\footnote{Because of the lower redshift observable for CIV, the same pixel size in angular units span a smaller physical size, of course.} While there is a general trend for the more luminous objects in Lyman $\alpha$ to be also bigger, interestingly the reverse seems to hold for the CIV bright sources (and potentially also for  OVI). Currently, we have no explanation for this anticorrelation in size vs. luminosity for the metal line bright objects.\\

\begin{figure*}
\includegraphics[width=168mm]{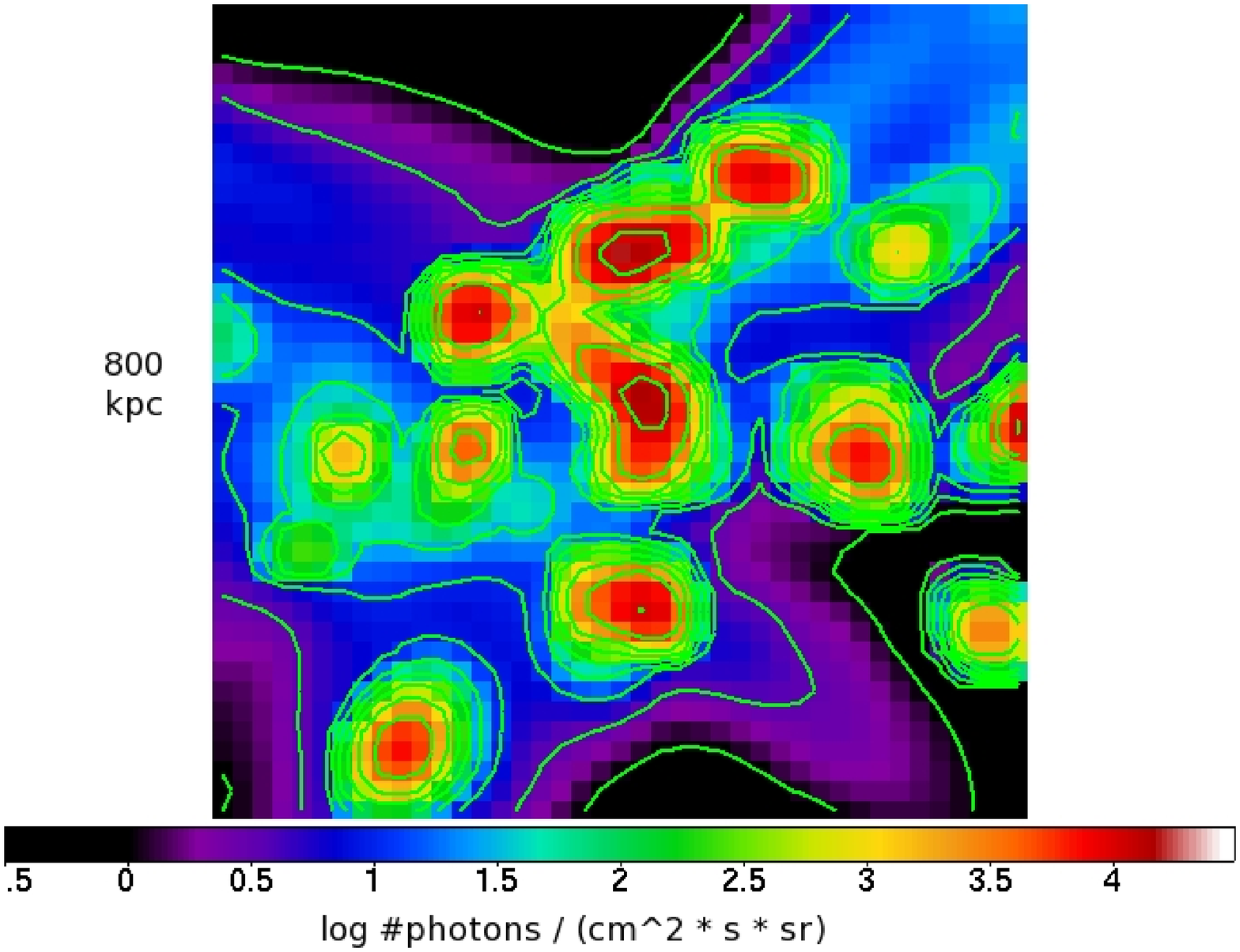}\caption{Example of a compact bright Ly$\alpha${} source at z=0.75 with a more complex spatial profile. The image has the same dimensions as Figure \ref{example_lya_075}. Between the numerous sources in this image, there are 'bridges' connecting the bright sources that can reach up to 1/10 of the peak emissivity.}\label{example_lya_1165}
\end{figure*}

\begin{figure*}
\includegraphics[width=168mm]{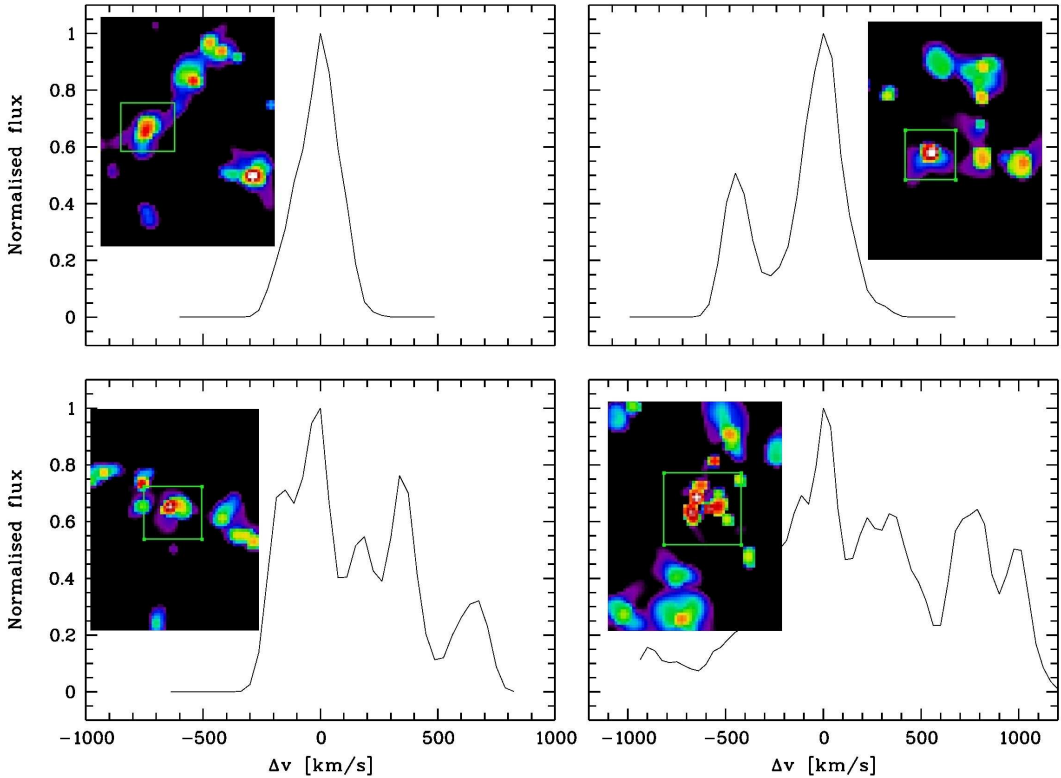}\caption{Examples of the 4 different classes of velocity profiles, here for the OVI emission. Note that the resolution of the representation chosen here (after resampling to the FIREBall specifics) is about 80 km/s per pixel. The coloured insets show the spatial images collapsed along the z axis as in Figure \ref{example_lya_075}. The velocity plots are derived by collapsing the area shown in green along the other two axis, and translating the differences in the z direction to velocities (for details see text). The {\bf upper left panel} shows a single peak in the spatial image, but the z-direction yields a profile that can be fitted well by a single Gaussian. The {\bf panel on the upper right} is an example of a similarly simply spatially structured source, but with a double peaked velocity profile. The {\bf lower left panel} exhibits again a single peak in x-y, but this time a much more complex, multiply peaked velocity profile. The {\bf lower right panel}, finally, has a complex spatial structure, in combination with a multiply peaked velocity profile. Note the slight change in scale for the velocity plot for this example. Both the CIV as well as the Ly$\alpha${} compact sources show velocity profiles very similar to the ones depicted here, but the interpretation for the Ly$\alpha${} emission is complicated by a variety of factors (see text for details).}\label{examples_of_profiles}
\end{figure*}

When we instead collapse the 3D cutouts from above along the two spatial axes perpendicular to the line of sight, we can obtain some information on the velocity structure of the emitting gas. Figure  \ref{examples_of_profiles} shows the profiles along the z direction for the a representative selection of bright sources in OVI along with their appearance in the spatial coordinates (green boxes).\footnote{The velocity profiles in CIV and Ly$\alpha${} show very similar structures, but we note that the interpretation of the Ly$\alpha${} structures is more complicated due to the variety of effects discussed earlier.} Here we plot the aggregated flux densities (integrated over the maxium source extent in x and y) versus the pixels' location in the cubes along z. If we assume the stretching in wavelength to be almost exclusively due to the peculiar velocities of the gas particles, we can translate the spectral extent into a velocity plot. Highlighted are four different scenarios into which we can group our objects. The zero velocity bin is defined arbitrarily to be the one with the highest flux, and we normalise by dividing through this peak flux. The top left panel shows the most common profile : a single peak (representative of about 65\%{} of the objects in Ly$\alpha${} and OVI, but less than half in CIV - see table \ref{source_properties}). If we fit these single peak profiles with a Gaussian, the FWHM ranges between 150 and about 400 km/s, with the bulk of them between 200 and 300 km/s.\footnote{Keep in mind that one pixel of 0.25 \AA{} represents already about 40 km/s. Hence, the lower limit is just a representation of our ability to resolve velocities.}{} The next largest group (25-30 \%{} for all transitions) exhibits double profiles, which themselves can come in at least three different flavours : the majority has two peaks of almost equal height, and a trough in between that never reaches zero flux (right upper panel). Some of the objects, however, have one peak rather dominant over the other one ($\geq$ a factor of 2 in maximum flux). These can be further divided into sources where there is either no gap of zero flux in between them, or where there is a separation (but less than 200 km/s). Note that for most of these objects the spatial structure is still indicative of one single object, as in this example, but for a few sources it is clear that there are two distinct features that are responsible for the double peak in the veolcity profile. A third category of objects has a multitude of different peaks, of varying heights with or without gaps of zero flux. While these represent less than 5\%{} for the Ly$\alpha${} and OVI emitting sources, they correspond to 10\%{} of the objects for the CIV emission. While the velocity spread is in most cases for all of the transitions confined to less than $\pm$400 km/s from the zero velocity bin, there are a few, rare cases where the whole structure can span almost 1300 km/s (see lower left and right panels). Those two examples again demonstrate that these complicated velocity profiles may be the result of many bright sources overlapping in the subcube extracted to derive the velocity profile (lower right), but need not be necessarily (lower left). While the former velocity profiles then can be readily explained by an overlap of a few single source profiles, the latter cases may be resulting either from multiple sources situated along the line of side, or be intrinsically more complex, possibly due to the imperfections of the extraction method (i.e. not taking into account spatial structures). Those latter cases, however, are only exhibited by a tiny minority of the sources ($<2\% ${} in all transitions).    \\

\begin{table*}\caption{Properties of typical bright emission regions as detected with {\it Duchamp}, at the redshifts specified in the text (Ly$\alpha$ z=0.76, OVI z=1.1, CIV z=0.35).}\label{source_properties}
\begin{tabular}{lccccc}
\hline
\hline
Transition & Number density   & Typical Maximum   & log L[erg/s] & Spatial extent  & Velocity structure \\
           & $\eta$[\#/(Mpc/h)$^3$]                 & Surface brightness & (median and range) & in FWHM [kpc proper] & \\
\hline
Ly-$\alpha$ & $38 \times 10^{-3}$  & $>3 \times 10^{4}$ CU & 41.8 (39..43.5) & majority : 50..150 kpc & Single peak 60\%, double peak 25\% \\
            &                     &                    & (lenient cuts)          &                        & Multiple peaks : 5\%,    \\         
            &                     &  & 40.9 (39..42.2) & extreme cases : & other : 10\% \\
            &                     &                    & (harsh cuts)          &  $>$ 200 kpc                    &  Maximum extent for   \\  
            &                      &  &                             &                             & multiple peaks : 1300 km/s \\
\hline
CIV & $25 \times 10^{-3}$  & $5 \times 10^{3}$ CU  & 38.5 (36..41)  & 25..100 kpc & Single peak 45\%, double peak 30\% \\
    &                      &                   & (cut independent) 
             &             & Multiple peaks : 13\%,           \\
    &                      &  &  &  & other : 12\% \\
    &                      &  &                             &                             & Maximum extent for \\
    &                      &  &                             &                             & multiple peaks : 800 km/s \\
\hline
OVI & $17.3 \times 10^{-3}$  & $1.5 \times 10^{4}$ CU & 39.2 (37..42) & 25..100 kpc & Single peak 60\%, double peak 30\% \\
    &                       &                      & (cut independent) 
             &             &   \\
    &                      &  &  &  & Multiple peaks : 2\%, other : 8\% \\
    &                      &  &                             &                             & Maximum extent for \\
    &                      &  &                             &                             & multiple peaks : 1000 km/s \\
\hline
\hline

\end{tabular}
\end{table*}

\subsection{Comparison with Ly$\alpha${} observations and origins of the bright Ly$\alpha${} emission}
 
  Our bright and compact sources, as displayed in Fig. 12 and with properties summarised in table \ref{source_properties}, reach  Ly$\alpha$  luminosities and space densities comparable to or larger than those obtained for the so-called Ly$\alpha$  emitters (LAEs) revealed by  a number of spectroscopic or narrow band imaging surveys at  high redshifts (e.g. Ouchi et al. 2008 and references therein). This comes as a surprise since these LAEs are identified as galaxies and their  Ly$\alpha$ emission essentially can be ascribed to star formation, a process not included in our emission models, but currently thought to be dominant in the  Ly$\alpha$  emission.  How is that possible and  why are the sources predicted by  the simulations not seen in current surveys? \\

Furlanetto et al. (2005)\footnote{Note however that Ly$\alpha$  emission from star formation is included in their analysis.} and Faucher-Gigu\`ere et al. (2010) have addressed these issues at  $z \sim 3$. Here, we compare the luminosity distribution  of the bright sources identified from our simulation with the only existing LAE survey at similarly low redshifts, the  GALEX spectroscopic surveys  at $0.20 < z < 0.35$ (FUV) and $0.65 < z < 1.25$ (UV)\citep[]{martin2005}. The GALEX surveys  have  found the LAEs to be less common and less luminous at the present time than at z $> 3$ \citep[]{deharveng2008, cowie2010}.  \citet[]{cowie2011} have shown  recently that most of this evolution occurs over the z = 0 - 1 range. Since the luminosity distribution of the sources depends more strongly on the self shielding correction (see also \citet{yang2006}) than on the redshift, we limit ourself to the case of a single redshift, z=0.35,  for which  the luminosity function of LAEs is less uncertain than at 0.75 \citep{cowie2011}. At a Ly$\alpha$  luminosity of about  3 $\times 10^{41}$ ergs s$^{-1}$, the lower limit for the LAE luminosity functions, the density of Duchamp sources (cut 7) is found to be 23 times larger than the LAE density of  3$\times 10^{-4}$ Mpc$^{-3}$ ($\Delta$logL=1)$^{-1}$ \citep{cowie2010}\footnote{Note that a comparison for cut 8 is less meaningful, since the curve is already declining at this luminosity value as a result of the method of analysis.}. At a Ly$\alpha$  luminosity of $\sim 10^{42}$ ergs s$^{-1}$, the excess factor decreases to 2 (for cut 7, but is as large as 100 for cut 8).\\

There are however several reasons for missing objects like our simulated sources with the slitless GALEX spectroscopic survey. First, a number of the sources from the simulation may not be associated to any continuum stellar light, whereas the extraction of GALEX spectra is based on the existence of a dispersed continuum. Second, in slitless spectroscopy line emission is smeared in the dispersion direction by the angular extent of the object. For a typical bright source size of 100 kpc  (i.e. 20 arcsec at z = 0.35) the  Ly$\alpha$ FWHM would be increased from 7  \AA\ (spectral resolution) to about   20  \AA , resulting in a loss of contrast over the continuum and, at low signal to noise ratio, a  non-identification as LAE or even a misidentification as AGN (if bright enough). Additionally, there is dilution in a direction perpendicular to dispersion, resulting in a loss of flux through the extraction window of 6 arcsec height.  Even given all the uncertainties, we do not consider resonant scattering to be able to affect significantly the Ly$\alpha$  photons from  cooling radiation, in contrast to the situation with Ly$\alpha$  photons resulting from star formation inside galaxies, where large optical paths through neutral gas may result in large quenching factors (if encountering dust particles) as well as extended halos (e.g. \citet{steidel2011}. In summary,  the characteristics of the existing GALEX slitless spectroscopic survey may be able to explain the non-detection of extra Ly$\alpha$  sources predicted by simulations. \\

    The fact that  star-forming particles are excluded from our simulations, combined with the angular extension of our Ly$\alpha$ Duchamp  sources, suggest naturally  an interpretation of these sources as CGM emission from Ly$\alpha$ cooling  radiation. This process has been repeatedly suggested to account for extended Ly$\alpha$ emission (e.g. \citet{haiman2000, fardal2001, dijkstra2006, dijkstra2009},   especially in the context of the so-called Ly$\alpha$  blobs (e.g. \citet{steidel2000, matsuda2004, saito2006}). This interpretation would be consistent  with our predictions of a larger density of sources with respect to that of LAEs,  since LAEs are themselves a small fraction of the star-forming galaxies (5 percent at $z \sim 0.3$, \citet{cowie2010}, 25 percent at $z \sim 3$, \citet{shapley2003}, for LAEs defined  by an equivalent width W(Ly$\alpha ) >$ 20  \AA ).\\

 The importance of Ly$\alpha$ cooling radiation has been challenged, however, by recent simulations \citep{zheng2010a}{} and observations \citep{steidel2011}, showing that spatially extended Ly$\alpha$ emission  is a generic feature of high-redshift star-forming galaxies, resulting from  Ly$\alpha$ photons initially produced by star formation and resonantly scattered by neutral atoms in the circumgalactic medium. Because of surface brightness threshold effects, the extended emission would hence be observed individually only in a limited number of cases. This would lead to an underestimation of the total  Ly$\alpha$ flux by an average factor of 5 (Steidel et al. 2011).  Such a factor, if applied to the luminosities of GALEX LAEs at $z \sim 0.3$, would be enough to make the resulting  Ly$\alpha$ luminosity function dominant over the luminosity distribution predicted from  Ly$\alpha$  cooling, at least for the cut 7 scenario (Fig. 11). In conclusion, this comparison validates our interpretation of the  Ly$\alpha$ bright sources extracted from our simulations as CGM emission, but leaves the contribution of cooling radiation to the total Ly$\alpha$ emission observed within the uncertainties of simulations, and therein especially of the self-shielding corrections \citep{faucher2010}.\\

\section{Extended Sources/WHIM}\label{sec:extended_sources}

This section gives a rough overview and some example plots for the structures that can be classified as 'filaments', i.e. here we take a look at the photons coming from regions outside the areas dubbed 'bright objects'. The main purpose is to be illustrative rather than quantitative and detailed, although some general features are summarised.\\

The procedure to produce the graphs below is to use the cubes resampled to FIREBall resolution (as described in 2.3), the specific cut values imposed are not important for these structures, the units of the data displayed here are in LU = \#photons/(sec * cm$^2$ * sr), as we integrate over the wavelength coordinate. The colour scaling of the images are chosen such that areas identified by the 3D-source finder {\it Duchamp} as 'bright sources' appear as white, compact 'blobs'.  The flux outside these sources represents only a tiny fraction of the total flux (depending on the specific cut used less than 5$\%$ in the best case for Ly$\alpha${},  and even less in the metal lines). In addition, much of the remaining flux when flagging out the pixels identified as belonging to compact sources, is contained in areas very close to the volumes cut out (.i.e. $<$ 50 kpc vicinity). Note that the contrast in flux density between the filaments and the bright blobs is up to 4 orders of magnitude, whereas there is another factor of $\sim$50 in the surface brightness delineating the filamentary structures from the overall 'average' emission level. \\

\begin{figure*}
\includegraphics[width=168mm]{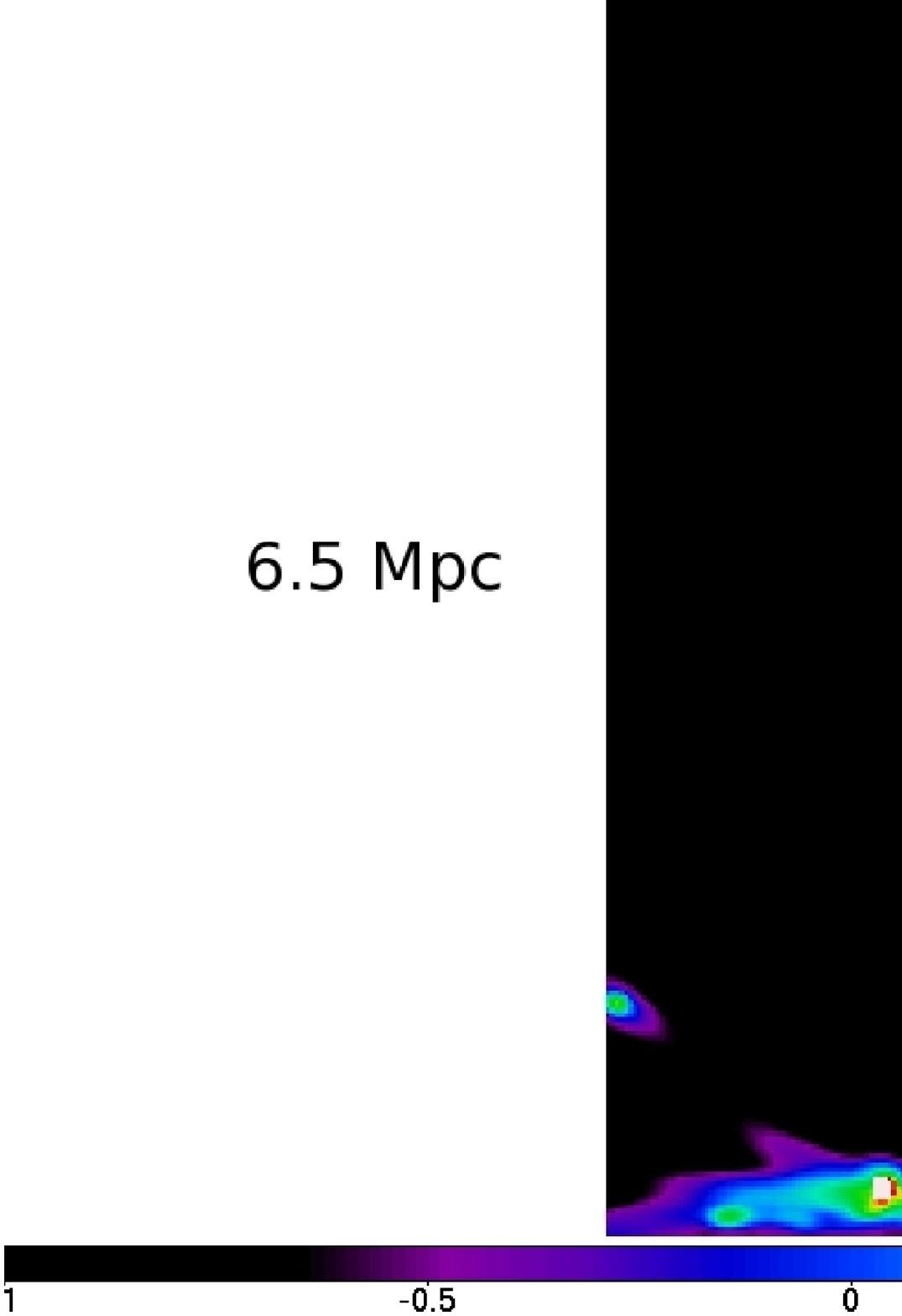}\caption{Example of a typical filamentary structure seen in Ly$\alpha${} emission at z=0.75. The size of the field in x and y is 900 x 900 arcsec$^2$, representing 6500 x 6500 kpc$^{2}$ (physical) at this redshift. The slice seen here has a thickness of 5.0 \AA, corresponding to about 6700 kpc in z direction. The bright, compact sources, which exhibit peak surface brightnesses of more than 4 orders of magnitude brighter than the filamentary structures connecting them, appear as white 'blobs' in the colour scheme chosen here. Note that the flux seen in the filaments is not spread uniformly, but strongly concentrated towards these bright sources, exhibiting a bright circum-galactic medium around these, before trailing off into 'bridges' forming the familiar sight of the Cosmic Web. Note that for Ly$\alpha$ an additonal contribution to the surface brightness may arise from the resonant scattering of Lyman line photons (a.k.a as 'photon pumping', see text for details.), which is {\bf not} included in this representation here. The physical conditions prevailing for the gas forming these large bridges are as follows : density n$= 5$ to $50 \times 10^{-6}$cm$^{-3}$, temperatures in the range of $\sim$50,000 $\leq T \leq \sim$120,000 K.}\label{filament_Lya}
\end{figure*}

Fig. \ref{filament_Lya}{} shows a typical situation for Ly$\alpha${} emission, as viewed by an instrument with roughly the specifics of FIREBall in angular and spectral resolution, but a much larger FOV (extended to 900 $\times${} 900 arcsec$^2$). The filamentary structure connecting the bright sources  has an extent in the x-y plane perpendicular to the line of sight of about 7200 kpc (physical), and extends over 6500 kpc in the z direction, assuming that velocity differences are not the main cause of the stretching over a wavelength range of 4.75 \AA.  Naturally, the boundaries of such structures are less well defined as the ones of the bright, isolated sources, so here we adopt the arbitrary definition whereby the outermost limit corresponds to the contour of the flux level 10 times higher than the median flux of all pixels. Then, the width of the long bridge, as measured perpendicular to the main direction between the bright sources, varies between 500 and 750 kpc. This rough 10:1 ratio of length to width is quite representative of the filamentary structures in Ly$\alpha$. An estimate by eye yields about 180 of such filaments in a cube of 25 $\times${} the FIREBall FOV, corresponding to a length per volume of 2.5 $\times 10^{-3}$ (Mpc/h)$^{-2}$, and a volume filling fraction of 0.1\%. It is interesting to compare these estimates with the values of \citep[]{bond2010}, who find and define filaments in their simulations via an algorithm that uses the eigenvectors of the Hessian matrix of the smoothed galaxy distribution. While their length measurements are of very similar size compared to ours (depending on the smoothing lengths they find $l/V = 1.9 \times 10^{-3}${} or $7.6 \times 10^{-4}$ (Mpc/h)$^{-2}$), their filling factors (combining this length estimate with their mean filament widths) are of the order of 5\%. This suggests that we are tracing similar structures (i.e. 'bridges' from and to massive and bright nodes of the Cosmic Web), explaining the similarity in overall length, but our focus on the HI emission outside of bound structures presumably results in much thinner sheets, as we are forced towards the densest parts of the gas. Following the definition above, the outermost regions of the filaments reach a surface brightness of about 0.1 CU, and even inside the filaments it usually remains below 5 CU, although very rarely bright spots (cf. the red-white areas of Figure \ref{filament_Lya}) can obtain up to 50 CUs. This is still a factor of at least 5 below the lowest brightness for the bright sources, but at least 3 orders of magnitude brighter than the median flux density of the whole cube. Typical physical conditions of the gas encountered inside such filaments are roughly as follows : the density ranges from n$= 5${} to $50 \times 10^{-6}$cm$^{-3}$, whereas the temperatures are between $\sim 50,000 \leq$ T $\leq\ \sim 120,000$K.\\ 

\begin{figure*}
\includegraphics[width=168mm]{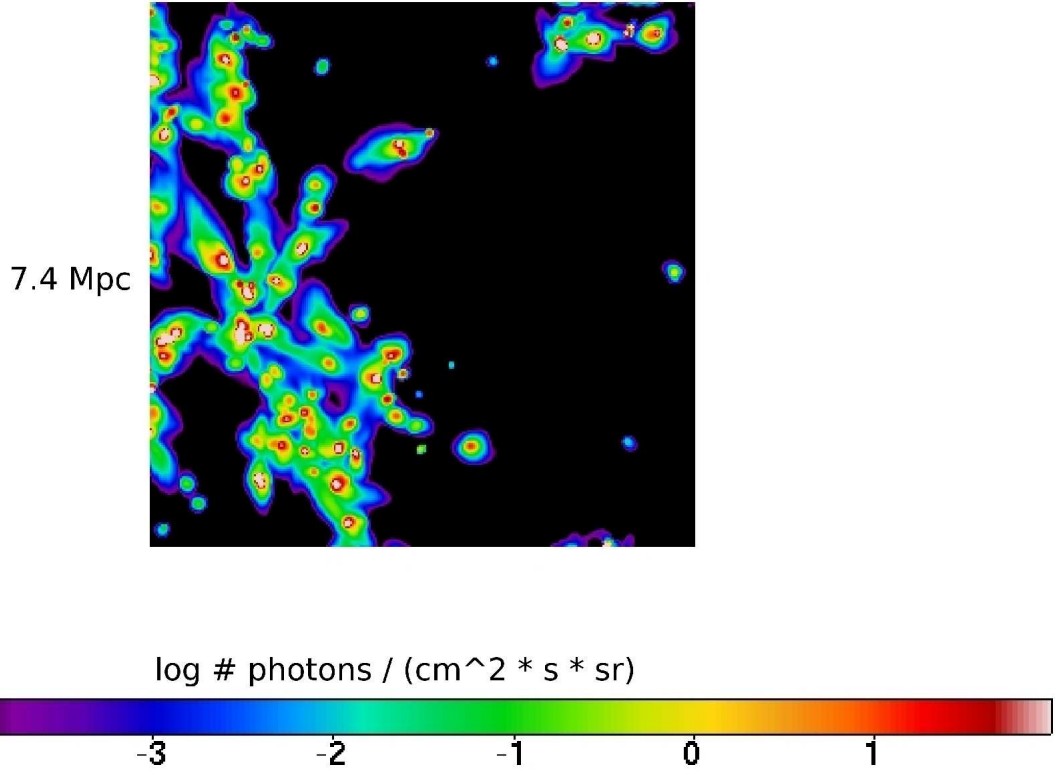}
\caption{Example of a typical 'filamentary' structure in OVI at z$_{em}$=1.1. The size of the field is again 900 x 900 arcsec$^2$, translating to $7.4 \times 7.4$ Mpc$^{2}$. The depth (5 \AA) represents 5.1 Mpc. Note the different scale in brightness as compared to Figure \ref{filament_Lya}  : the filaments connecting the bright sources are now about three orders of magnitude fainter. The physical conditions of the gas forming the filamentary structures are very similar to the ones seen in Ly$\alpha${}  : the metallicity of such gas ranges from $10^{-3.5}$ up to $10^{-2.5}$, which is the main factor in those to be less bright. Note, however, that the flux is concentrated towards the bright sources even more than is the case for Ly$\alpha$.}\label{filament_OVI}
\end{figure*}

\begin{figure*}
\includegraphics[width=168mm]{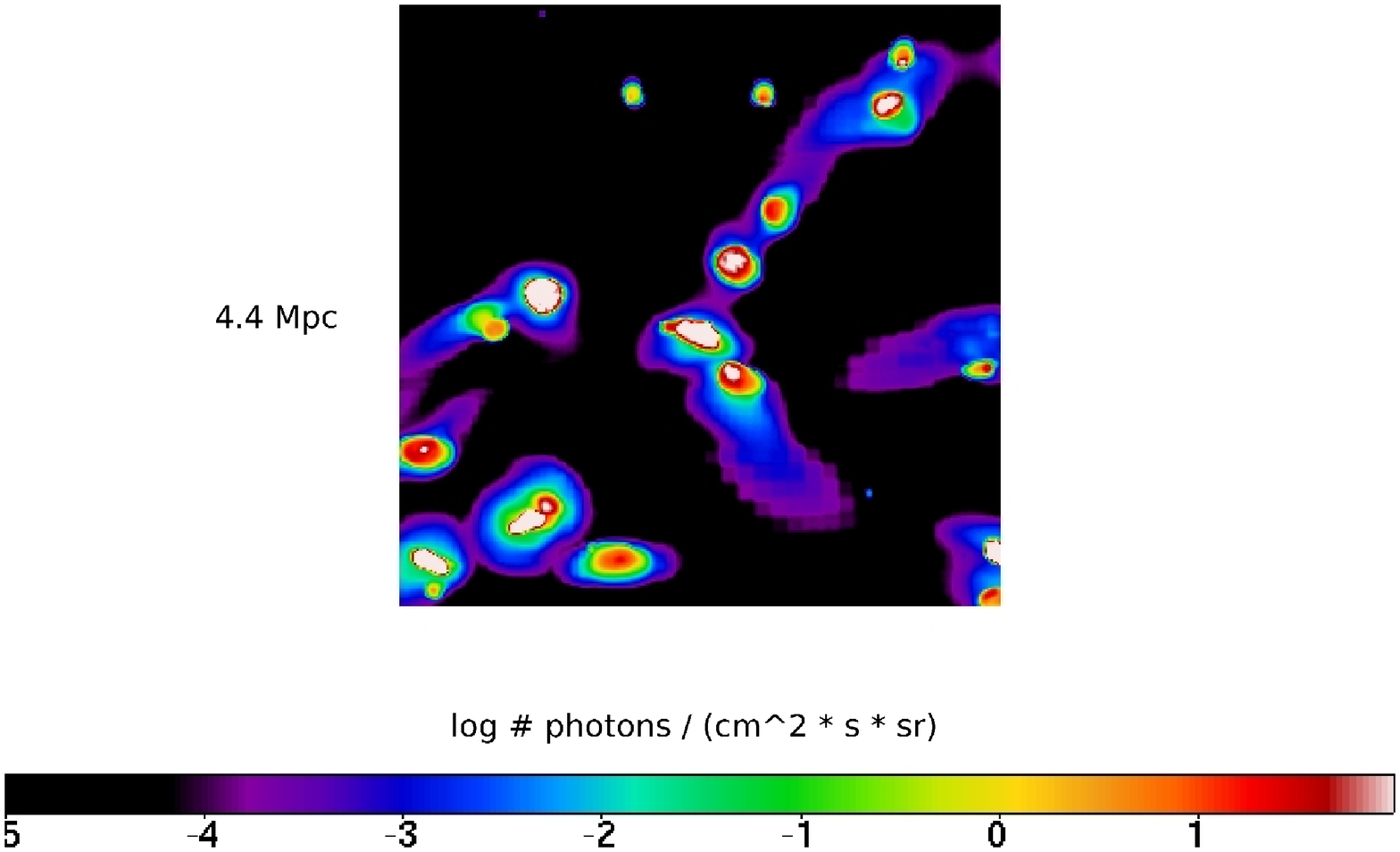}\caption{Example of a typical 'filamentary' structure in CIV at z=0.35. The size of the field is again 900 x 900 arcsec$^2$, equaling $4.4 \times 4.4$ Mpc$^{2}$, while the depth is 8.4 Mpc. The bright nodes are still connected by filamentary structures (with roughly the same physical conditions as in the OVI and Ly$\alpha${} cases).}\label{filament_CIV}
\end{figure*}

Similar filaments can be seen in the metal line transitions OVI and CIV. These still consist of 'bridges' connecting two nodes (see Figures \ref{filament_OVI}{} and \ref{filament_CIV} for examples in OVI and CIV), however are much fainter than their Ly$\alpha${} counterparts. In addition, the steeper decline of the emissivities in the metal lines with both temperature and especially density, leads to the remaining bright spots outside the compact sources to be more (OVI) respectively {\it much} more (CIV) concentrated around these sources than for Ly$\alpha$. By comparing the estimated gas emissivities in Ly$\alpha$, OVI, and CIV for this parameter space (see Figures \ref{lya_pie_emissivity} to \ref{civ_pie_emissivity}), and noting that they are all around the same value (assuming solar metallicity), we can conclude that the main reason for the metal line transition filaments to be so much fainter, is the gas metallicity, which tends to be slightly below $10^{-3}$ of the solar value, but can reach up to $10^{-2}$ in a few, isolated spots, usually near the bright sources, where it is reasonably to expect the metal enrichment of the IGM to take place first.\\

\section{Observing Strategies}\label{sec:observing_strategies}

 The idea of observing the emission from the IGM may be traced back to the early investigations of  the nature of the UV (non-ionizing) background (e.g. \citet{davidsen1974, paresce1980}{} and the idea that the resonance lines of HI and HeII from the IGM may  contribute to this background.  The role of clumping was already recognized as primordial for the possibility of detection \citep{paresce1980b}. Since then, the association between the IGM emission and the UV domain  has come to the forefront  with the predictions of the WHIM and the CGM \citep{cen1999, dave2001}.  The development of this phase is expected to be  maximum at low redshift and  its  temperature implies observations  only through Ly$\alpha${} and high ionization species (OVI, CIV, etc) at UV rest-frame wavelengths.  Except for the Ly$\alpha$\ fluorescence observations  which make sense at high redshifts and can be attempted from the ground \citep{rauch2008, cantalupo2005},  the WHIM and CGM-oriented  observations  have therefore to rely upon space or balloon experiments in the UV domain.

\subsection{Detectability of bright, compact sources}

\begin{figure*}
\includegraphics[angle=270,width=168mm]{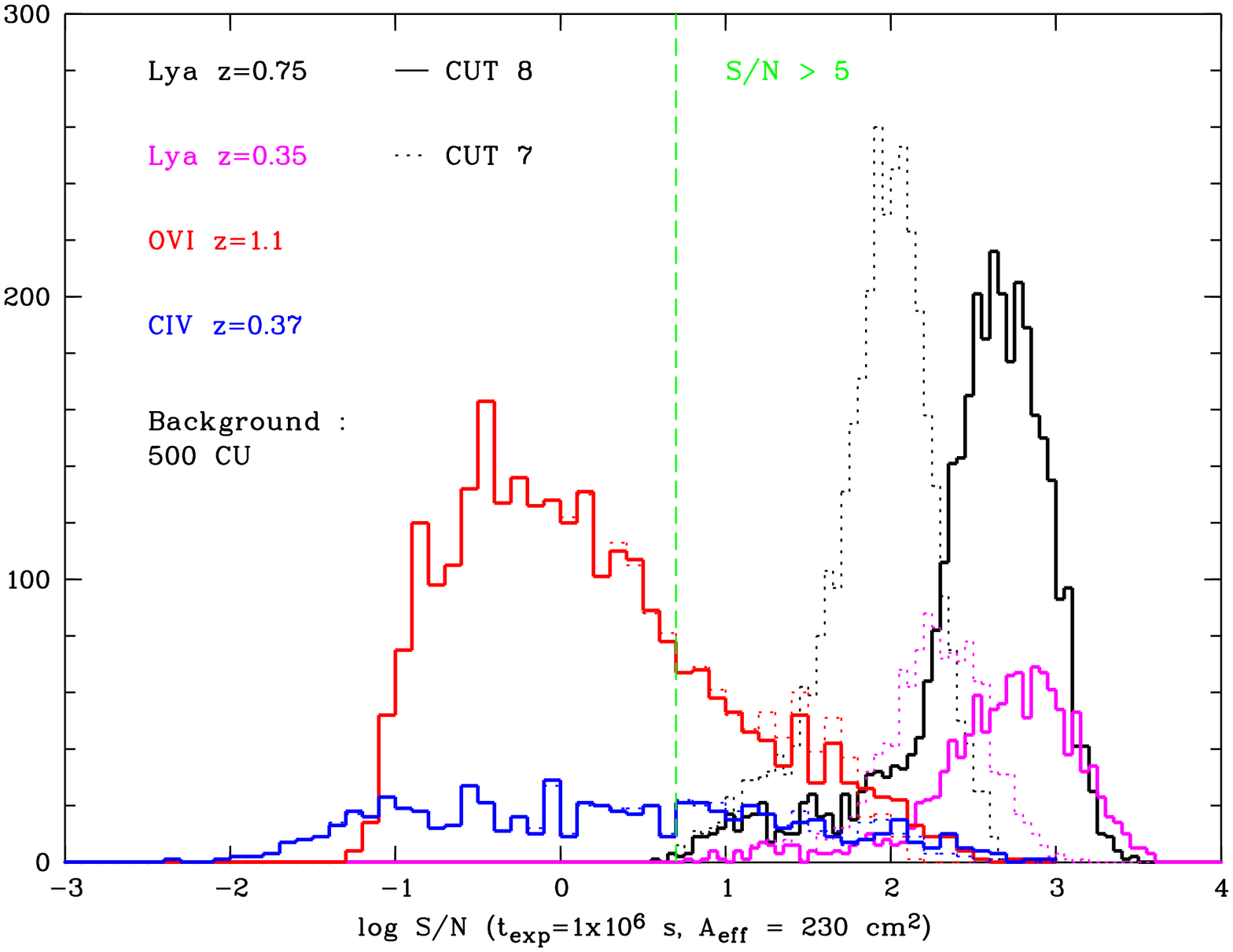}\caption{The distribution of the signal-to-noise ratios expected for the objects found in the data cubes, as detailed in the text and shown in Figure \ref{luminosity_distributions}. Here we are assuming an exposure time of 10$^6${} seconds, and an effective collecting area for an instrument of 230 cm$^2$, typical for a proposed deep field observation in future space-based missions. The noise is assumed to be a combination of photon noise and the extra-galactic background, which we take as 500 CU. Note that the absolute number of objects is a function of the FOV and the spectral coverage. The latter has been chosen to be representative of FIREBall, and may be raised by a factor of 5 or more for future space-based missions. The former is in this case, a $3 \times 15 \times 15$arcmin$^{2}$ area of the sky. In addition to mass resolution effects (as is the main contributor for Ly$\alpha$), the shape of the distributions for the metals is also influenced by the imperfections of the source extraction algorithm, especially for OVI towards the low S/N end. For details, see the text.}\label{SN_texp1e6}
\end{figure*}

What are the prospects of detecting the compact, bright sources identified via {\it Duchamp} as detailed in section 4.2. with current technology ? And what may be the best instrumental approach set out for this task ? Trying to gauge the detectability in a first approach, we have translated the {\it Duchamp}{} results into signal-to-noise-ratios (S/N) for each source, based upon the following simplifying, yet conservative assumptions. Note that certain aspects of the source detection already rely upon some fairly general instrumental characteristics (i.e. spatial and spectral resolutions). Furthermore, we assume to be dominated by the (cosmic) background as a noise source, and take this to be at a uniform level of about 500 CU \citep[]{brown2000, morrissey2005, murthy2010}. For an effective telescope area of 230 cm$^{2}$ (representative of a meter-class space-based UV-experiment) and an exposure time of $1.0 \times 10^{6}$s (representative of a typical deep exposure), we obtain the S/N ratios depicted in Figure \ref{SN_texp1e6}{} for the three different transitions at three different representative redshifts. The histograms have been computed from 3 emission-rich sightlines for each transition, totaling a combined 'observed' Field-of-View in each case of 3 $\times 15 \times 15${} arcmin$^{2}$. \\

As can be easily seen, in this scenario we would be able to recover {\bf all}{} of the Ly$\alpha${} emitting sources, regardless of the specific cuts (solid and dotted black lines) or redshift (black or cyan curves) with high confidence (the vertical green line indicates a S/N of 5.0). Hence, even uncertainties in the precise source position and its extent, will not jeopardise detection at all - note that we have implicitly assumed to know both quantities perfectly in this approach for our simulations (degrading this precision by e.g. including random areas classified by {\it Duchamp} as not belonging to the compact sources only marginally affects the S/N for any sensible number of additional pixels, as we have tested). Thus, for Ly$\alpha$, the two main problems to solve are : a. guaranteeing a large enough FOV to allow for enough detected sources (e.g. to perform meaningful statistical analyses), and b. finding suitable tracers to guide target selection  from observables established as good proxies to emission from the simulation. The latter point is, of course, only relevant for instruments with multi-object spectrographs, as sources with S/N $>$5 will be detectable by blind searches with integral-field spectrographs.\\

The situation is somewhat different for the metal line transitions : both for the OVI and the CIV emitting regions, only a fraction of our simulated sources will be detected, regardless of the specific cut used for the self-shielding, which is largely unimportant for the brightness of these systems, as detailed earlier. As the blue  and red histograms in Fig. \ref{SN_texp1e6}{} show, about half of the CIV (at a redshift of 0.37) , and 2/3 of the OVI sources (at z=1.1) in our scenario may be missed. This fraction changes towards lower redshifts for OVI specifically, but note that the absolute number of sources per FOV then drops dramatically because of the enormously smaller area covered by the same opening angle (this effect can be seen e.g. in the two different redshifts for the Ly$\alpha${} sources). Hence, an additional complication for targeting metal lines emission areas compared to the search for Ly$\alpha${} sources is that one furthermore needs to be able to select tracers that guarentee to pick up the {\bf bright} end of the distribution. Or be prepared to mostly rely upon stacking approaches because the majority of sources may individually lie below a detection limit, but collectively gather enough photons to statistically infer average quantities. \\

Under the given assumptions the S/N scales with $\sqrt{t_{exp} \times A_{eff}}$. Hence for different combinations of exposure time and telescope size, the distribution simply needs to be shifted horizontally by the appropriate factor. Furthermore, the absolute source number is a linear function of the FOV, and hence the distribution can be shifted up- and downwards depending on this quantity, rendering this graphical representation a versatile tool to judge mission goals.\\

A typical exposure time for a balloon-based experiment like e.g. FIREBall is about two orders of magnitude lower than indicated here - meaning that one can expect FIREBall to be successful in detecting Ly$\alpha$, and eventually a very small number of the brightest metal line emitting areas, depending on the size of the FOV. Note also, that in the case for Ly$\alpha$, we have specifically excluded photons stemming from star-forming regions, which may indeed be the major contribution to the flux, even at larger distances from the galaxies themselves due to the resonant nature of the Ly$\alpha${} transition (e.g. see Steidel et al. 2011 for examples at high z). While this additional contribution will facilitate detection of bright areas, it also complicates the analyses as the different components contributing to the total flux need to be separated. \\

Also note that we have thus far relied upon a fixed spectral coverage of only about 200 \AA, comparable to the capabilities of FIREBall. Extending this coverage by almost a factor of 10, as planned for certain space-based missions, will not only vastly increase the survey volume and hence the number of detectable sources - in this context it is worth reminding ourselves that we do not find a significant evolution in the Ly$\alpha${} space density over the redshift intervals probed here, and thus the number counts should scale almost linearly with the survey volume - but also allow for the simultaneous observation of all three transitions, possible yielding important insights into the emitting material's physical state. \\

\subsection{Detection of filamentary structures}

As expected,  a quick glance at the PDF for Ly$\alpha${} emission (Figure \ref{lya_pdf_volume} for Ly$\alpha${} at z=0.75, but also valid for all other redshifts) confirms that by far the largest fraction of the cosmic volume lies below any realistic chance of a direct detection, as the peak of the distribution reaches mere levels of $\sim 10^{-3}${} CU. Furthermore, a closer inspection of the flux remaining outside of the bright sources, which represent only 0.5 to 5\%{} of the total flux to begin with, reveals that is is not homogeneously distributed, but rather shows a complex spatial structure in our simulation, with isolated spots reaching up to 50 CU over volumes as small as 5 voxels.\\

We have conducted a few tests on a stacking approach. First, we have isolated by eye a few of the brighter filamentary structures such as seen for example in Figure \ref{filament_Lya}, and added all their flux after removing the bright sources within them. Secondly, we have gone even one step further in our 'naive' idealisation, and simply combined voxels outside the bright sources in descending order of brightness {\it regardless} of their actual position, until we maximised the S/N for the combined flux in that fashion (note that there is a maximum number of voxels after which adding new cells in fact leads to a diminished S/N due to the additional voxel being so dominated by noise rather than signal). Both methods, applied to a mock observation with an exposure time of 10$^6${} s and an effective telescope area of 230 cm$^2$, lead to marginal detections of the aggregated voxels, irrelevant of redshift and the specific cuts for self-shielded gas. Again, recognising that these two scenarios are extremely optimistic or even overly idealistic - in a real observation e.g. there is no a-priori information where exactly the filaments themselves or their brightest part are located - it is obvious that dectecting the faint, filamentary structure even just in Ly$\alpha$\footnote{As we have seen in paragraph \ref{sec:extended_sources}{}, there are indeed filamentary structures in the metal lines very similar to the ones observed for Ly$\alpha$, yet due to the metallicity being much lower than unity, these are so faint that we can safely focus on the - as we have shown - already difficult to detect emission from hydrogen alone.}{} will remain a challenging endeavour with current technology. \\

In reality, however, at least two aspects  that may change this pessimistic outlook should be mentioned. First of all, as we have seen earlier, the mass resolution of the simulation may lead to an underestimate of the gas emission, specifically because it depends so non-linearly on temperature and densities. While this is most certainly the case for the brightest regions in the immediate vicinity of the densest and/or hottest areas, it seems less likely that large effects should be expected  for the only slightly overdense and colder structures constituting the filaments, but we caution that even here clumping on scales smaller than resolved may occur.\\

Secondly, it is worth comparing the emission calculated with our simulation, based upon photo- and collisional ionization in an isotropic radiation field consisting purely of the average extra-galactic background, with estimates for at least one other process that we have identified already in 2.4 as not being incorporated properly into our modeling, i.e. the photon-pumping of local non ionizing continuum photons from galaxies. In order to calculate the potential additional emission at the order-of-magnitude level, we have devised the following toy-model at $z$ = 0.35: we have assumed all filaments to be dust-free uniform cylinders illuminated by the UV continuum of young galaxies uniformly spread along their central lines. A 300 kpc proper FWHM diameter in HI has been taken, in rough agreement with the structures we have seen in paragraph 2.2. The cumulative comoving filament length per cubic megaparsec has been estimated by eye in one of our 900 $\times$900 arcsec$^{2}$ cones to be at $\sim$0.006 Mpc$^{-2}$. In the present coarse approximation, effects due to filament inclination on the line of sight would induce changes at the few tens percent level and have been neglected. Those assumptions result in an average column density along a filament radius of N$_{HI} \sim$14.0 cm$^{-2}${}, able to scatter 0.4 \AA~ from a central continuum source when b = 30 km/s, independently of the velocity distribution in the gas. The total luminosity scattered by HI in the filaments has then been taken equal to the total energy absorbed along all lines of sight, with the central source's luminosity chosen to produce the same luminosity density near 1216 \AA~ as measured by GALEX in the FUV \citep[]{schiminovich2005}. We thus assume that most of the continuum near 1216 \AA\ in a given portion of space is produced by flat continuum galaxies distributed in the filaments, a reasonable assumption at moderate redshifts where the UV flux is dominated by galaxies of small mass. We find the HI Ly-$\alpha$ luminosity density "pumped" under those assumptions to be 4 times above that from our simulation. The average filament surface brightness is found to be of the order of 10 LU, suggesting that areas with surface brightness in the hundred LU might not be uncommon given the luminosity spread of the illuminating galaxies in the real world, and the presence of clustering as already pointed out by e.g. \citet{kollmeier2010}. This coarse computation brings the emission from such a filamentary structure into the realm of achievable sensitivities for a space-mission, provided the FOV is large enough to cover a significant area. Keeping in mind that velocity information may not be as crucial (or interesting) for this gas  (cf. e.g. figure \ref{filament_Lya}, that shows the filaments not being spread over more than 100 to 200 km/s in velocity space), and also that high-spatial resolution is not mandatory, leaves us with the possibility of relaxing some of the instrument characteristics in order to maximise the S/N to be expected from the extended, faint emission regions. \\
 
We caution, however, that while such an additional contribution to the Ly$\alpha${} radiation can boost the chances of observability, it will tell us little about the physical state of the gas other than the distribution of neutral hydrogen and its velocity field (in conjunction with information about the non-ionising UV flux emanating from galaxies). The latter information though is relevant to the infall of cold gas.


\section{Summary}\label{sec:conclusions}
Exploiting a state-of-the-art large-scale AMR structure simulation, we examine the possibility of observing UV line emission at redshifts $0.3 \leq z \leq 1.2$, focusing on the Ly$\alpha$ and two metal line transitions (OVI and CIV). The emissivity estimates for the gas in the simulation utilises the spectral synthesis code CLOUDY, adopting options most suitable for the treatment of the situation at hand (i.e. assuming a \citet{haardt2001} UV background with a zero percent escape fraction from galaxies as incident background at the simulation redshift ; turning off the line pumping for all transitions). Being interested in the emission from extra- and circumgalactic gas, we exclude all photons coming from cells in the simulation that are starforming. Furthermore, we are bracketing the emission of self-shielding gas by setting it either radically to zero emissivity, or by assuming collisional ionisation equilibrium at the temperature and density provided by the simulation.\\

Generally speaking, we encounter two types of emission regions : relatively compact, bright, isolated objects, and large, faint filamentary structures permeating all of the volume, connecting such dense knots. The former represent (presumably) circumgalactic material that is of high enough density and temperature in order to emit at surface brightnesses that can easily be reached with current instrumentation, while the latter will certainly remain challenging even for future satellite missions, as detailed below.  By degrading the AMR simulation to the spatial and spectral resolution of FIREBall, an instrument with characteristics very typical for UV balloon observations, we can assess realistically the chances of observing the gas, and furthermore can describe in detail which physical characteristics may be extracted from such observations.\\

Applying {\it Duchamp}, a standard source extractor algorithm, to the observationally-minded datacubes, we are able to isolate the bright, compact objects for all three transitions at the three different redshifts dictated by the FIREBall spectral window (1990 - 2260 \AA, resulting in 0.64 $<$z$_{em} < 0.84${} for Ly$\alpha$, 0.93 $<$z$_{em} < 1.19${} for OVI, and 0.28 $<$z$_{em} < 0.46${} for CIV). While extending over less than 0.1\%{} of the cubes' volumes, these bright sources carry more than 95\% of the flux, even for the least optimistic assumption for the treatment of the self-shielding gas. The number density for the Lyman $\alpha${} bright sources in our simulation is $\eta (Lya, z=0.8) = 38 \times 10^{-3}$ (Mpc/h)$^{-3}$, whereas the densities for bright sources in CIV and OVI are $\eta (CIV, z=0.37) =  24.8 \times 10^{-3}$ (Mpc/h)$^{-3}$, and $\eta (OVI, z=1.1) = 17.3 \times 10^{-3}$ (Mpc/h)$^{-3}$.\footnote{The number of sources, however, is strongly dependent on the simulation resolution. See the Appendix for details regarding convergence.} While for the brightest areas in Ly$\alpha${} the range depending on the specifics of the self-shielding treatment can be up to two orders of magnitude, the effect for the metal lines is minimal, as the main contribution to the light emitted stems in their cases from gas of temperatures well above any of the self-shielding cuts. Hence, our estimate for the median luminosity for the Ly$\alpha${} sources ranges from log L [erg/s] = 40.9 to log L = 41.8, while the median luminosities for the metal lines are robust against changes in the treatment of self-shielding : log L (OVI) = 39.2, and log L (CIV) = 38.5. Keep in mind, however, that while the luminosity distribution for Ly$\alpha${} sources peaks sharply roughly at its median value, the distributions in OVI and CIV are much broader, without a clearly preferred peak. The overwhelming majority of all sources, Ly$\alpha${} and metal lines, extend over roughly spherically symmetrical areas between 50 to 100 kpc (proper) FWHM (when fitting images collapsed along the velocity component/z direction of the cubes by 2D Gaussians), but in a few percent of the cases, the source can obtain sizes of up to 300 kpc. The morphologies of such large sources are often more complex, hinting at their being several, individually unresolved smaller ones in  a group or cluster. Interpreting the sources' profile along the z axis of the cube as a velocity profile, we encounter a variety of different cases : on the order of 60\%{} of the objects (again in all three transitions) exhibit a single peak, with a median FWHM of 215 km/s, and a range from 150 $< \sigma < 400$ km/s. Next, there is a large group of sources ($\sim$25\%{} of the cases) which show a double peaked profile, roughly split in half by ones that are symmetrical and others where one peak clearly dominates. Interestingly, for this group of objects with double-peaked velocity profiles their spatial profile looks in almost all cases still indicative of one single source, which is not true for the last group of objects with more complex, multiply peaked velocity structures. While some of the members of this category can have relatively simple spatial structure, many of them are clearly examples of multiple, yet unresolved individual sources within the window over which we extract the velocity information. In general, the velocity spread over the whole source is less than $\pm$400 km/s from the brightest pixel, but in rare cases can span almost 1300 km/s.\\

In contrast to these bright, isolated sources with sharp 'edges', the morphology of the filamentary structures connecting the nodes of the Cosmic Web are more complex with much less well defined boundaries. For all possible scenarios dealing with the self-shielding cuts, they collectively carry less than 5\%{} of the flux in Ly$\alpha$, and even less in the metal lines. Adopting a straight-forward definition to delineate these from the surrounding medium (outermost contour has to have a surface brightness 10 times larger than the median flux of the whole cube), we extract filaments that exhibit roughly a 10:1 ratio of width to length. The width, defined as length perpendicular to the longest segment connecting bright nodes, of these structures are between 500 and 750 kpc (physical). Within the definition above, the boundaries for the Ly$\alpha${} filaments occur roughly at 0.1 CU, and remain inside those usually below 5 CU, although rare bright spots may obtain up to 50 CU. The brightness of filaments in the metal transitions is about 2 orders of magnitude below that of Ly$\alpha$, largely an effect of the metallicity inside the filaments being of that order of magnitude below solar. Typical physical conditions of the gas encountered inside such filaments are roughly as follows : density ranges from n$= 5${} to $50 \times 10^{-6}$cm$^{-3}$, temperatures are between $\sim 50,000 \leq$ T $\leq\ \sim 120,000$K, and gas metallicity tends to be slightly below $10^{-3}$ of the solar value, but can reach up to $10^{-2}$ in a few, isolated spots, usually near the bright sources. The flux distribution inside the filaments is not at all homogeneous, with about 50\%{} of it stemming from areas in the close vicinity of the bright sources ($\leq \sim$ 50 kpc). Given all these characteristics of the filamentary structures it is clear that their detection will remain a challenging task for the near future. \\

Applying simplifying and conservative assumptions about source extraction as well as background and detector noise, we conclude that the bright sources in Ly$\alpha${} associated with the CGM will easily be picked up with high confidence by instrumentation with current (or soon to be deployed) technology. For example, {\it all}{} of the simulated Ly$\alpha${} sources at z=0.76, and a large fraction of the metal line transitions OVI and CIV exhibit S/N ratios in excess of 10 when 'observed' by a meter-class space-based UV telescope with an effective area of 230 cm$^{2}$ in a moderately long exposure of 10$^{6}$ seconds. While a ballon-based experiment with shorter duration observations will also be able to retrieve the majority of the Ly$\alpha${} objects, the fraction of objects bright enough in the metal lines drops to the few percent level. A spatial resolution of 3-5 arcseconds suffices to minimally resolve the bright CGM areas extending up to 200 kpc, and is currently easily obtainable technologically. Furthermore, a spectral resolution of $\sim$0.5 \AA{} allows to extract the velocity profiles of these objects, and hence obtain kinematic information on the light-emitting gas distribution. The space density of the bright sources is high enough that even with a relatively short spectral coverage of $\sim${} 200 \AA, already feasible with current detectors and dwarfed by possible space-based instruments that are poised to cover much more of the UV, surface densities can be achieved that guarantee sufficient numbers of objects in a survey field of a few arcmin$^{2}$, such that meaningful population analyses can be performed. Given these scenarios, the next question to be addressed has to be how to select good tracers in a possible surveying campaign. Note that we have purposefully neglected all stellar light emanating from the galaxies themselves, or even scattered Ly$\alpha${} radiation, and hence our surface brightness estimates are strictly lower limits. In fact, distentangling such different components from the radiation we have simulated in our approach will become an important caveat for any survey. Hence, simply choosing UV-bright galaxies as targets may not be the best way forward. Other source characteristics (e.g. metallicity, or environment) may serve better, and we are currently investigating this question.               
     
\section*{Acknowledgments}
This work has been funded within the BINGO! ('history of Baryons: INtergalactic medium/Galaxies cO-evolution') project by the Agence Nationale de la Recherche (ANR) under the allocation ANR-08-BLAN-0316-01. Access to the HPC resources of CINES was granted and funded under the allocation 2010-x2010042191 made by GENCI (Grand Equipement National de Calcul Intensif).


\bibliographystyle{mn2e}
\bibliography{references_bingo_paper_I}



\section*{Appendix : Convergence tests}

Especially towards areas of high density and/or temperature, the ability to resolve small enough scales to capture many of the relevant processes at least to some detail is crucial. As pointed out in the main text, e.g. the number of the bright sources is a strong function of the simulation (mass and spatial) resolution. A variety of tests can be performed in order to assess whether a simulation has converged onto stable values in a a wide range of parameter space. Because we are mainly interested here in the light emission of the gas outside startforming particles, we focus in this Appendix on two aspects regarding the bright, compact sources, and the surface brightness of all voxels outside these.

\begin{figure*}
\includegraphics[angle=270,width=168mm]{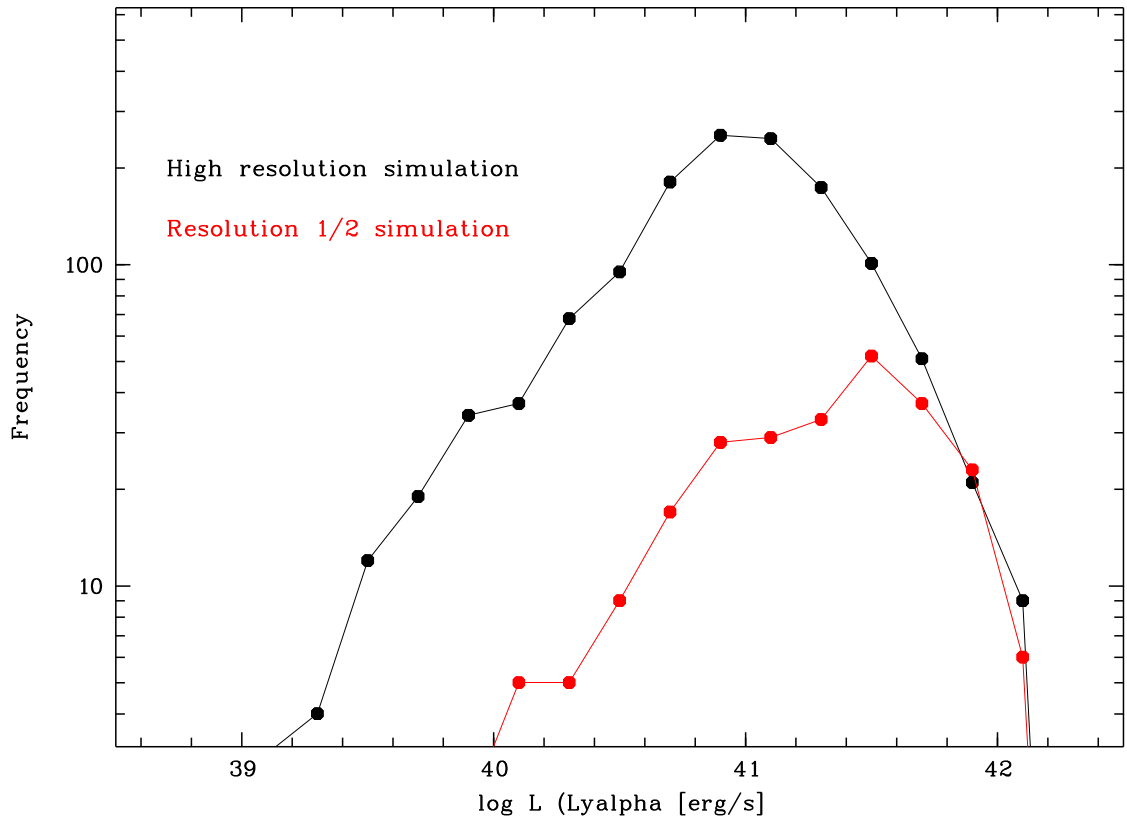}\caption{The luminosity distribution of the bright Ly$\alpha${} sources at z=0.75 extracted with the same source finding parameters in two cubes of the same volume with different spatial and mass resolutions. The black curve shows the results obtained in the simulation used throughout the paper, whereas the red histogram is derived from a simulation with a spatial resolution of a factor of 2 lower (in each dimension), and hence a mass resolution of a factor 8  worse. Note the strong dependence of the {\it number} of sources on the resolution : increasing the resolution in this fashion leads to about 5 times more sources in the same volume. The majority of the new sources, however, are much less luminous, and hence their contribution to the total light is small. The two curves start matching around log L $\sim$ 41.6, and hence we conclude that down to this luminosity the simulation has converged.}\label{convergence_bright_sources}
\end{figure*}

Fig. \ref{convergence_bright_sources}{} shows the luminosity distribution of
bright sources in Ly$\alpha${} isolated in exactly the same manner for two simulations with different resolution. Both simulations are examined at the same redshift (z=0.75), and over exactly the same volume. Whereas the black histogram shows the luminosities of detected sources in the original high resolution simulation, when applying the most lenient approach for handling the self-shielding, the red curve depicts sources found in a simulation that was run with half the resolution in each spatial direction, and hence with a mass resolution of a factor of 8 less. Clearly, the {\it number}{} of such bright sources is a strong function of the resolution : there are about 5 times as many such bright sources in the high resolution simulation. However, it is also obvious that almost all of the 'new' sources (as compared to the lower resolution output) are of lower luminosity, and their combined emission does not add substantial light to the total Ly$\alpha${} radiation, which is dominated by the most luminous sources. Above an approximate limit of log L (Ly$\alpha$) $\sim$ 41.6 both curves start to match. Hence, we conclude that the simulation has converged down to that limit of luminosity. In this sense, the number densities for the bright sources are strictly only lower limits.\\

\begin{figure*}
\includegraphics[angle=270,width=168mm]{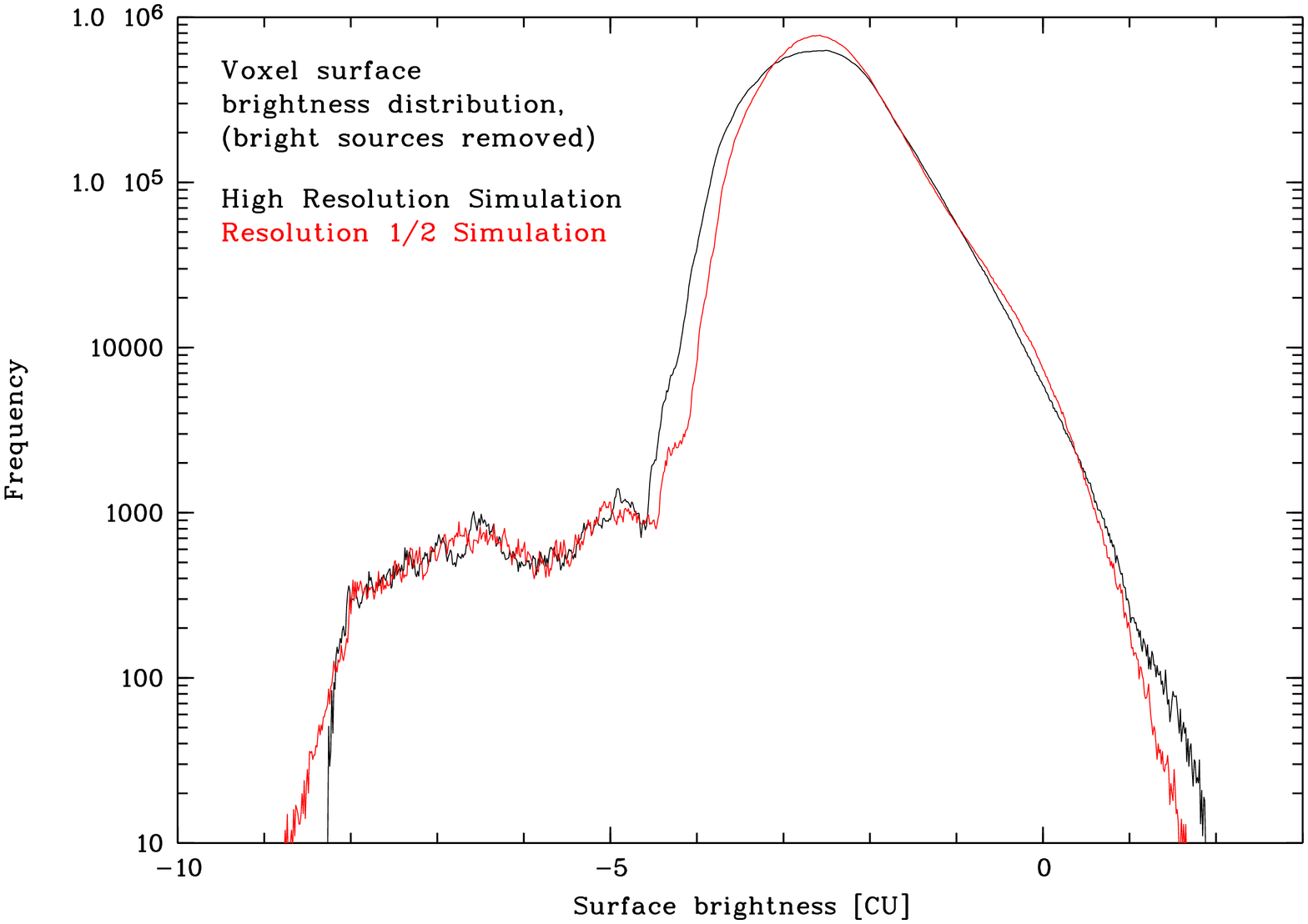}\caption{The surface brightness distributions of two Ly$\alpha${} cubes at z=0.75 after removing the bright sources. The black curve shows the results obtained in the simulation used throughout the paper, whereas the red histogram is derived from a simulation with a spatial and mass resolution of a factor 8 worse, as in Fig. \ref{convergence_bright_sources}. Unlike for the bright sources, there is little difference in the two histograms, indicating - as expected - that for areas of moderate overdensity the simulation has reached convergence. There is a small number of pixels inside filamentary structures that exhibit surface brightnesses of about half a dex higher in the higher resolution cube, but those are so rare and spread out over wide volumes that they do not contribute significantly to the brightness of one specific filament.}\label{convergence_filaments}
\end{figure*}

Outside of the bright sources, however, the situation is different. Fig. \ref{convergence_filaments} shows the surface brightness distributions of the same two simulation outputs mentioned above, after in each one the voxels identified to be belonging to bright compact sources (and a cushion of 5 pixels around these) have been removed. Those two histograms are remarkably similar, with only a very tiny minority of areas exhibiting a slightly brighter flux in the high-resolution cube. In that regard, we are able to assess that for the mildy over- to underdense regions singled out by cutting out the bright sources the simulation, not very surprisingly, has converged, rendering the predictions for those areas stable.\\

Note that for both metal lines similar arguments hold.  


\bsp

\label{lastpage}

\end{document}